\documentclass[sigconf,nonacm]{acmart}

\usepackage{gensymb}
\usepackage{array}
\usepackage{dblfloatfix}    
\usepackage[absolute,overlay]{textpos}

\AtBeginDocument{%
  \providecommand\BibTeX{{%
    \normalfont B\kern-0.5em{\scshape i\kern-0.25em b}\kern-0.8em\TeX}}}

\begin{document}

\thispagestyle{empty}

\title[TCBs: Thermoformed 3D Printed Circuit Boards]{Thermoformed Circuit Boards: Fabrication of highly conductive freeform 3D printed circuit boards with heat bending}


\author
{Freddie Hong}
\email{t.hong19@imperial.ac.uk}
\orcid{https://orcid.org/0000-0002-6318-9723}
\affiliation{
  \institution{Imperial College London}
  \streetaddress{Imperial College Road}
  \city{London}
  \state{London}
  \postcode{SW7 1AL}
  }

\author
{Connor Myant}
\email{connor.myant@imperial.ac.uk}
\orcid{https://orcid.org/0000-0003-4705-4009}
\affiliation{
  \institution{Imperial College London}
  \streetaddress{Imperial College Road}
  \city{London}
  \state{London}
  \postcode{SW7 1AL}
}

\author
{David Boyle}
\email{david.boyle@imperial.ac.uk}
\orcid{}
\affiliation{
  \institution{Imperial College London}
  \streetaddress{Imperial College Road}
  \city{London}
  \state{London}
  \postcode{SW7 1AL}
}


\begin{abstract}
{\parskip=0pt
Fabricating 3D printed electronics using desktop printers has become more accessible with recent developments in conductive thermoplastic filaments. Because of their high resistance and difficulties in printing traces in vertical directions, most applications are restricted to capacitive sensing. In this paper, we introduce Thermoformed Circuit Board (TCB), a novel approach that employs the thermoformability of the 3D printed plastics to construct various double-sided, rigid and highly conductive freeform circuit boards that can withstand high current applications through copper electroplating. To illustrate the capability of the TCB, we showcase a range of examples with various shapes, electrical characteristics and interaction mechanisms. We also demonstrate a new design tool extension to an existing CAD environment that allows users to parametrically draw the substrate and conductive trace, and export 3D printable files. TCB is an inexpensive and highly accessible fabrication technique intended to broaden HCI researcher participation.}
\end{abstract}
\keywords{
3D printed electronics, conductive filament, hybrid additive manufacturing}

\begin{teaserfigure}
  \includegraphics[width=\textwidth]{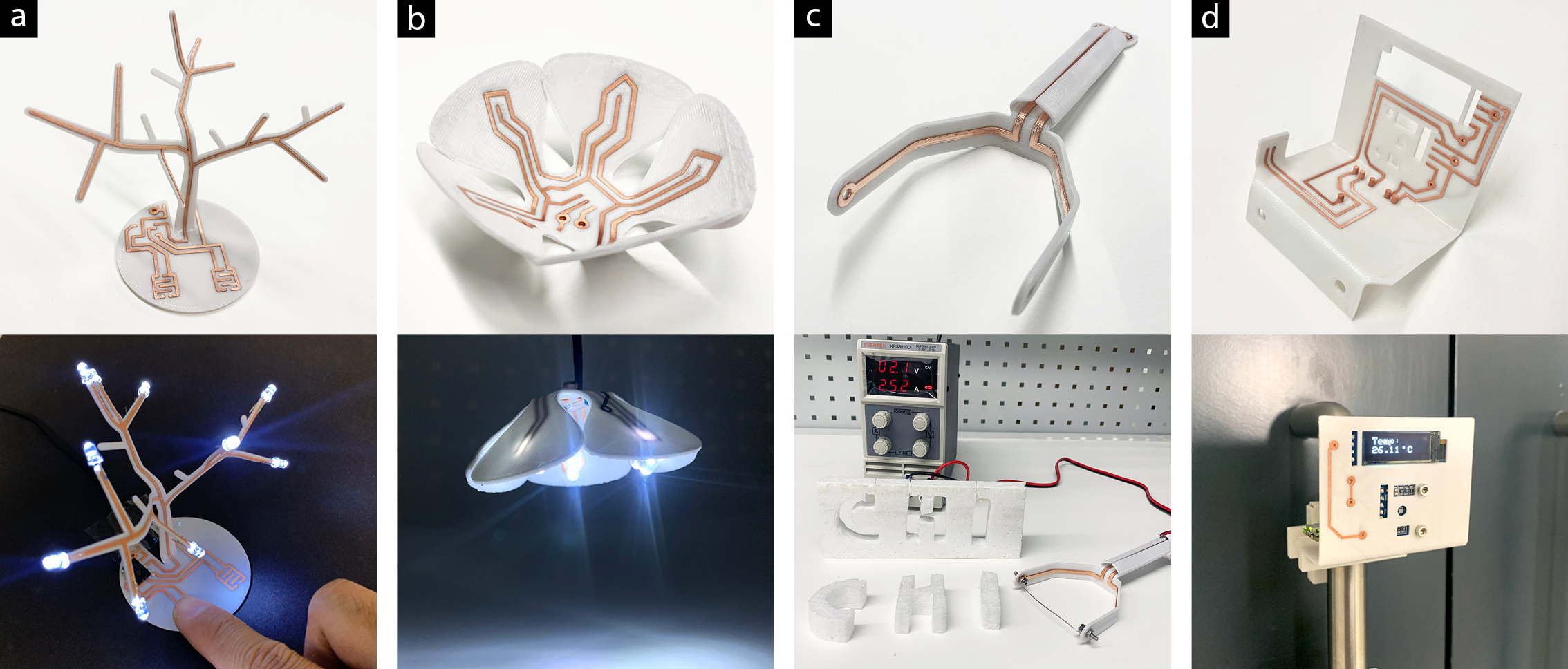}
  \caption{Examples of Thermoformed Circuit Board Devices: a) Tree-shaped touch sensitive lamp with 555 timer IC; b) Light sensitive pendant light; c) Hot wire foam cutter; d) Stand-alone contactless temperature sensor with OLED display}
  \Description{range of examples using thermoformed circuit boards are shown. 1a shows a double sided circuit board thermoformed into shape of a tree attached to 555 IC cheap to turn light on and off using touch. 1b shows a pendent light thermoformed into shape of a flower which is sensitive to the light using LDR sensor. 1c is a hot wire cutter with 2.5A current. As a demonstration the shape 'CHI' is cut out from the Styrofoam block. 1d is a contactless thermometer attached to a door handle}
  \label{fig:teaser}
\end{teaserfigure}
\maketitle

\section{Introduction}

Recent advances in additive manufacturing technologies have enabled rapid prototyping and integration of complex and highly customized digital designs into physical objects with minimal manufacturing processes. Additive manufacturing can respond quickly to urgent socio-economic challenges, such as those presented by COVID-19, by rapid prototyping functional products like face shields and hands-free adaptors~\cite{Choong2020ThePandemic,Tino2020COVID-19Medicine}. Under a traditional manufacturing process, this may have taken too long to produce and deliver. There has been emerging interest in building interactivity into 3D printed objects, which has thus led to various developments in 3D printable conductive materials and hybrid additive manufacturing processes allowing users to construct electrically functional objects with bespoke form factors. These advances in 3D printed electronics promise a range of new opportunities including increased freedom of design \cite{10.1145/3334480.3383174,10.1145/3313831.3376617,10.1145/3173574.3174015,Wu20153D-printedSensors} and reduction in time and cost of manufacturing bespoke prototypes \cite{10.1145/3290605.3300797,10.1145/3290605.3300858,10.1145/3379337.3415898,Wang2018Plain2Fun:Circuits}, all of which contribute to the next generation of device design and expanded prosumerism.

Within additive manufacturing research, various approaches have been proposed. These include extrusion of conductive ink via direct writing \cite{Lopes2012,Wasserfall2015} and droplet-based printing \cite{OptomecElectronics}, extrusion of carbon nanotubes  \cite{Vatani2015ConformalSensors,Vatani2015CombinedSensors}, ultra-sonic embedding of copper wires \cite{Jr2018HybridFabrication,Kim2017}, and extrusion of conductive thermoplastic filament via fused deposition modelling (FDM) \cite{Flowers20173DFilament}. There has also been commercial development of a desktop 3D electronics printer called Voxel8 \cite{Voxel8Specifications}, which combines FDM printing of thermoplastic with the extrusion of conductive silver ink. These technologies remain largely inaccessible within maker communities, design studios and human computer interaction (HCI) research because of the high barriers to entry in terms of cost for capable machines and limited design capability owing to poor conductivity of low cost materials. 

Although the introduction of Voxel8 and copper electroplating conductive PLA techniques~\cite{Kim2019,Lazarus2019,Angel2018,Vaneckova2020CopperElectrodes} have shown promise, constructing conductive elements beyond the XY plane remains a challenging task. This is due to the inherent characteristics of planar construction, where material is extruded layer upon layer causing weaker or broken material bonding along the Z-axis. This in turn causes poor conductivity for vertical interconnects in comparison to horizontal interconnects, and limits the possible design diversity of 3D printed electronics beyond the XY plane. Various HCI researchers have, therefore, taken alternative approaches to construct interactive 3D objects through adhesion of conductive copper tape and films~\cite{Umetani2017SurfCuit:Prints,10.1145/3290605.3300858,Savage2012Midas:Objects,Roquet20163DCircuits,10.1145/2807442.2807511}, and attachment of 2D inkjet printed circuits~\cite{10.1145/3173574.3174015,Olberding2015Foldio:Electronics}. These methods require extensive manual effort in addition to being limited in applicability due to their restriction to single-sided circuits.

\textit{CurveBoard}~\cite{10.1145/3313831.3376617} is a recent method to construct 3D breadboards to prototype electronic functions using conductive silicone filled inside 3D printed housing. In exchange for its re-configurability, however, it has limited design resolution due to required thickness and spacing of the conductive channels, making it best suitable for early stage prototyping. 

Aiming to expand capabilities for 3D printed electronics, while also establishing an immediately employable technique, we present a novel method to construct rigid, freeform, conformal, double-sided 3D printed electronic circuit boards. This is achieved by heat bending 3D printed thermoplastic conductive traces, and extends to withstand high-current applications through a process of electroplating. Since Thermoformed Circuit Boards (TCB) can be manufactured with common Polylactic acid (PLA) filament and FDM 3D printers, this technique is highly accessible for HCI researchers. With TCB, designers can explore new possibilities and create interactive 3D objects incorporating electronics in expressive form factors that are \textit{only} manufacturable by 3D printing. We summarize the contributions of this paper as follows: 

\begin{itemize}
    \item We demonstrate of novel fabrication technique of thermoforming 3D printed conductive traces to construct freeform, rigid and double-sided circuit boards using inexpensive and accessible equipment and materials
    \item We experimentally evaluate electrical and mechanical properties of TCBs
    \item We provide a parametric design editor integrated with the 3D modelling environment to simplify the specification and layout of circuit elements, including traces, vias and sockets suited for 3D printing
    \item We demonstrate the capabilities of TCB by prototyping example applications with various form factors, electrical characteristics and interaction mechanisms
    \item We openly share our research material online, including our modified 3D printer slicer settings, Rhino and Grasshopper files, materials and suppliers list, and ‘how-to’ guides\footnote{\url{https://github.com/FreddieHong19/Thermoformed-Circuit-Board}}
    
\end{itemize}

\section{Related Work}

Our research spans the fields of freeform electronics, hybrid additive manufacturing and digital fabrication in HCI.

\subsection*{Fabrication of Electronic Circuits on 3D Surfaces} Several methods of fabricating small and complex circuits conformal to 3D surfaces have been explored. For example, Aerosol Jet system \cite{OptomecElectronics} by Optomec uses droplet-based deposition of electronic ink (diameter: 1 to 5~$\micro$m) on multi-axis system to print complex conformal circuits on 3D substrates. Adams et al. \cite{Adams2011} have employed extrusion of conductive silver ink on 3-axis positioning stage to construct electrically small antenna on hemispheric substrates. Molded Interconnect Device (MID)~\cite{3DLDS}, which uses laser structuring on molded plastic to electroless plate conductive traces on the 3D surface, is also a possibility. MID techniques are often employed for high volume consumer parts, such as those found in cars and smartphones. Despite high performance, these approaches are difficult for general users to adopt. Aerosol Jet systems are too expensive for individuals and small studios to own, and extrusion of conductive silver ink on 3D surfaces requires complex control of its rheological property and extrusion angle. Employing MID device is also mostly infeasible as the manufacturing process is suited for very large quantity production, and the flexibility for design development is restricted due the to molding requirement.

Alternatively, various ‘user in the loop’ approaches have been proposed. For example, SurfCuit \cite{Umetani2017SurfCuit:Prints} and Midas \cite{Savage2012Midas:Objects} use manual placement of copper tapes on 3D objects. Saada et al. \cite{Saada2017HydroprintingStructures} and ObjectSkin \cite{10.1145/3161165} hydroprint circuits onto curved surfaces using conductive silver ink printed on polyvinal alcohol (PVA) film. Since the copper tape is commercially available and inexpensive, and conductive silver ink can be printed using consumer standard inkjet printer. Both approaches are very easily accessible for a wide range of potential users. However, these approaches can only fabricate single-sided circuits, which have limited applicability due to restrictions on circuit complexities, in addition to the manual labor involved in their fabrication processes. Our approach of 3D printing conductive trace and the substrate together in a single step followed by thermoforming can more conveniently fabricate double-sided freeform circuits with less manual effort. We thus offer a new method of fabricating conformal electronics which extends from the recent survey on the application of 3D conformal electronic presented by Huang et al.~\cite{Huang2019AssemblySurfaces}. 

\subsection*{Fabricating Interactive Object from Conductive PLA} There has been a series of commercial developments of  conductive thermoplastic filaments that are directly usable on FDM printers. Since FDM is the most popular form of inexpensive and low entry 3D printing, the introduction of conductive PLA can be seen as an ideal approach for many potential users wanting to prototype electronic devices. However, as the resistance of the commercial conductive PLA is too high, most of their applications have been limited to capacitive sensors (Capricate \cite{Schmitz2015Capricate:Objects}, MonoTouch \cite{Takada2016MonoTouch:Gestures},  ./Trilaterate \cite{10.1145/3290605.3300684}, PrintPut \cite{10.1007/978-3-319-22701-6_25}). Most recently, \textit{ProtoSpray} \cite{10.1145/3334480.3383174}explored the use of conductive PLA as a base electrode for electroluminescence and created freeform displays on arbitrary surfaces. While \textit{ProtoSpray} has shown great potential, the method stands to benefit from having access to more uniform, conductive and integrated electrical traces on freeform surfaces.

Copper filled conductive filament, e.g. \textit{Electrifi} by Multi3D \cite{Multi3D}, has improved the resistance of the conductive PLA and has showcased expanded applications including antennas and inductors~\cite{Flowers20173DFilament}. However, the conductivity of the printed trace remains inadequate for circuits with actuators. In addition, the current methods of 3D printing conductive trace using layer upon layer construction have very limited design freedom due to inconsistent quality of the printed trace for sloped angles. For instance, for \textit{Electrifi}, the resistance of the printed trace for 90\degree vertical is 7 times higher than the resistance of the printed trace in the horizontal direction. For sloped angles lower than 45\degree, printed traces can be broken into separated segments because of the staircase effect caused by planar construction.
As a method to increase the conductivity of the printed trace, Angel et al. \cite{Angel2018} and Kim et al. \cite{Kim2019} have electroplated the printed traces with copper. This method shows a significant change to the conductivity of the printed trace and has also shown the ability to withstand high current. Inconsistent quality of the printed trace for inclining surfaces can still, however, cause uneven finishes in plating. In this paper, we explore the technique of thermoforming the flat printed parts to construct a freeform surface that maintains the parallel structure of printed layers even for arbitrary folds. We next explore the electroplating of thermoformed printed parts to expand currently limited application of conductive PLA. Our investigation in changes of the resistance prior to thermoforming, post thermoforming and post copper electroplating is discussed in Section 3.

\subsection*{Construction of 3D Interactive Objects from 2D Materials} Folding and inserting 2D printed circuits into 3D shapes have been the most popular methods to rapid prototype 3D interactive objects. However, due to lack of rigidity of the papercraft, the applications of the folded paper circuits are flimsy (e.g. Foldio \cite{Olberding2015Foldio:Electronics}, PrintGami \cite{Roquet20163DCircuits}). Recently, FoldTronic \cite{10.1145/3290605.3300858} employed the honeycomb structure to exploit the foldability of sheet materials to construct structural devices, but this inherently limits freedom for component placement. Although not electrical, LaserOrigami \cite{Mueller2013LaserOrigami:Objects} also explores thermoformability of acrylic sheets to lasercut, bend and stretch the 2D material to construct functional 3D objects. Similarly, our approach starts by printing flat circuit boards, but since plastics retain shape after thermoforming, the final objects are structurally rigid and offer higher degrees of design freedom.


\section{Thermoformed Circuit Board}
\subsection*{Motivation and Inspiration}
The main objective of this work is to establish a new fabrication technique that can be immediately used in practice. For many individual makers, FDM 3D printer may be the only viable option to wholly manufacture electrically interactive 3D objects. The popular method of employing printable conductive silver ink among additive manufacturing researchers is out of reach for most individual users wanting to fabricate 3D electronics. Furthermore, embedding conductive elements inside the geometry causes many problems. These include: i) In multi-material printing, stringing and oozing can occur as the nozzle travels above the printing part. For electronic fabrication this leads to cross-contamination and short circuits, which are difficult to detect and fix when embedded; ii) Lengthy 3D printing times and multiple layer changes accrue high failure rates for prints; iii) Inconsistencies in mechanical and electrical properties in prints due to the staircase effect.

From the designer's perspective, we wanted to create a technique with the following benefits: i) Quick printing times with low failure rates; ii) Inexpensive and accessible in any DIY environment; iii) Increased design freedom; iv) Highly conductive; v) Increased automation. We drew inspiration from the look and feel of MID devices and recently introduced electroplating conductive PLA techniques to construct 3D circuit boards that share similar functional advantages with MID devices, but are manufactured with desktop machines.

\subsection*{Concept}
\begin{figure}[h]
  \centering
  \includegraphics[width=1\linewidth]{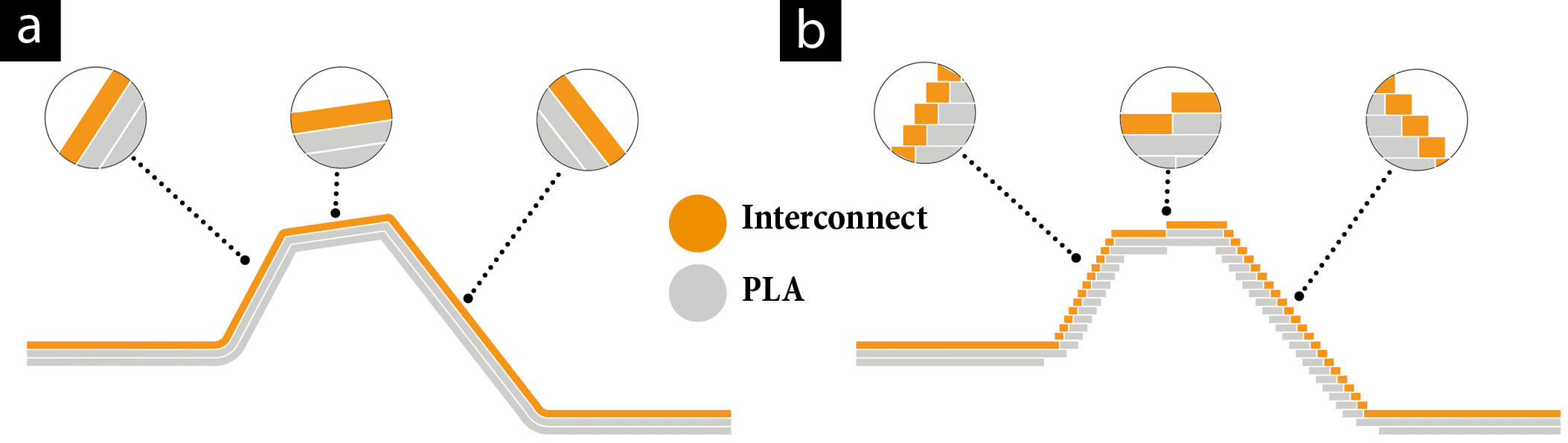}
  \caption{Thermoformed print structure (a); Planar construction print structure (b)}
  \label{fig:PrintStructure}
  \Description{Comparative illustration showing the layer structures of TCB against conventional 3D  printing method for sloped angles, left shows thermoformed method of constructing 3D trace and right shows planar construction method of constructing 3D trace. TCB layer shows smooth and continuous layer structure whereas conventionally 3D printed structure shows stair stepping effect.}
\end{figure}

TCB is constructed using two thermoplastic materials: PLA, an insulating substrate, and copper-based conductive PLA for electrical traces. TCB exploits the low glass transition temperature of the PLA (T\textsubscript{g}~< 60{\degree}C) to bend the printed parts using a hot air blower. The main benefit of thermoforming flat 3D printed parts into freeform surfaces over 3D printing the end shape is that it can construct 3D parallel traces for inclining surfaces with far greater continuity in the quality, and thus better conductivity, of the printed traces (Figure \ref{fig:PrintStructure}). 

\begin{figure}[h]
  \centering
  \includegraphics[width=1\linewidth]{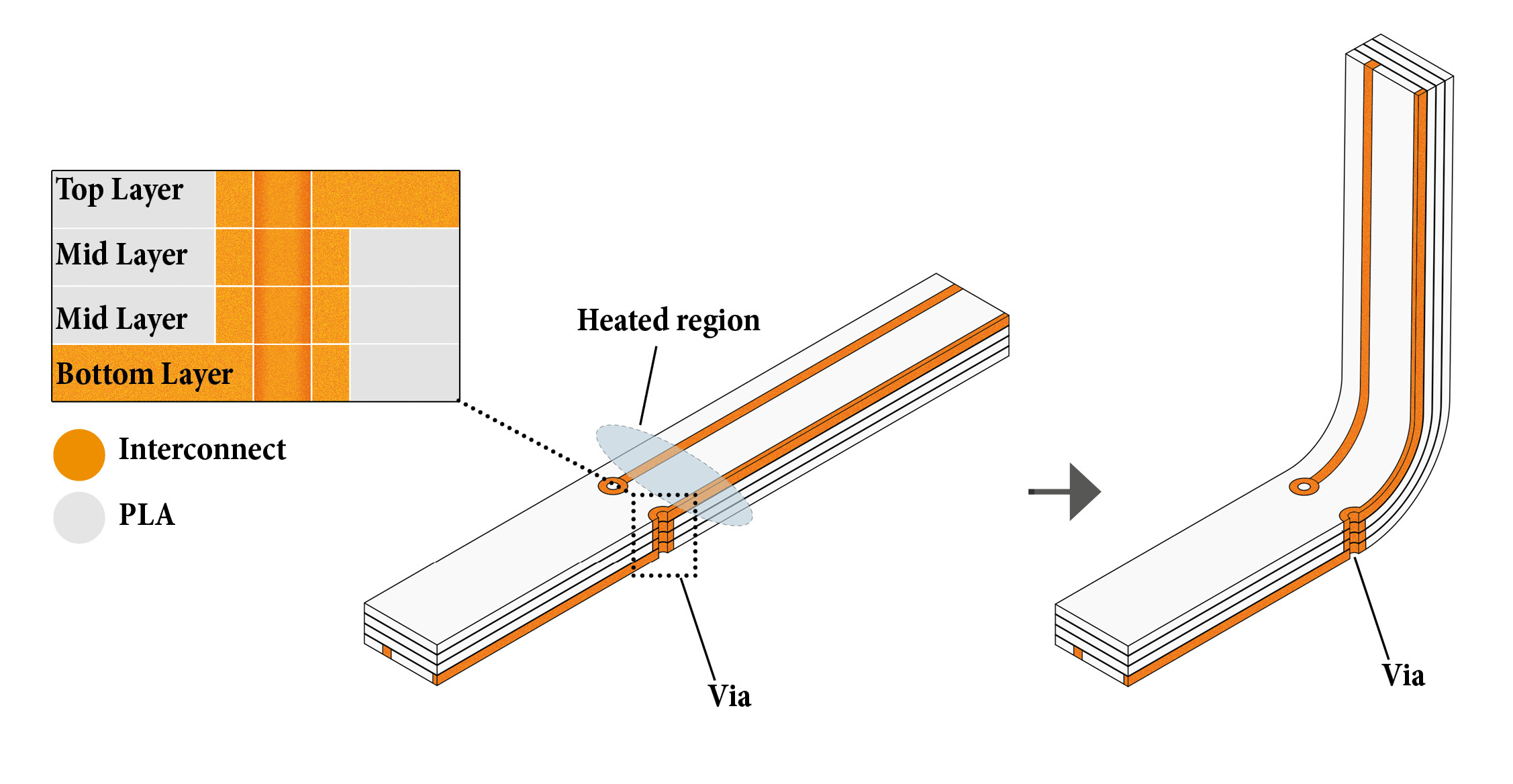}
  \caption{Layering of the TCB (left), TCB as printed (middle) and thermoformed board (right)}
  \Description{Two illustration of the TCB layers structure showing flat printed TCB and 90$\degree$ heat bent TCB. TCB's conductive trace maintain its continutiy for the ange change}
  \label{fig:BendingConcept}
\end{figure}

Typically, double-sided TCBs consist of 4 printing layers: one bottom layer, two mid layers and one top layer. The mid layers provide good electrical insulation between the top and bottom circuits due to strong dielectric properties of PLA~\cite{Dichtl2017DielectricAcid}. Top and bottom layer circuits are connected by through-hole vias (Figure \ref{fig:BendingConcept}). TCB is a highly accessible fabrication technique that only requires a single digital fabrication step, where the additional steps can be achieved using widely available and inexpensive equipment. Furthermore, since TCB only requires a minimal amount of printing layers and short printing time, we have used a single extruder FDM printer with an improvised tool changing G-code to manually swap the filaments. Finally, heating the target region of the printed part is performed to thermoform the 3D shape. 

\begin{figure}[h]
  \centering
  \includegraphics[width=1\linewidth]{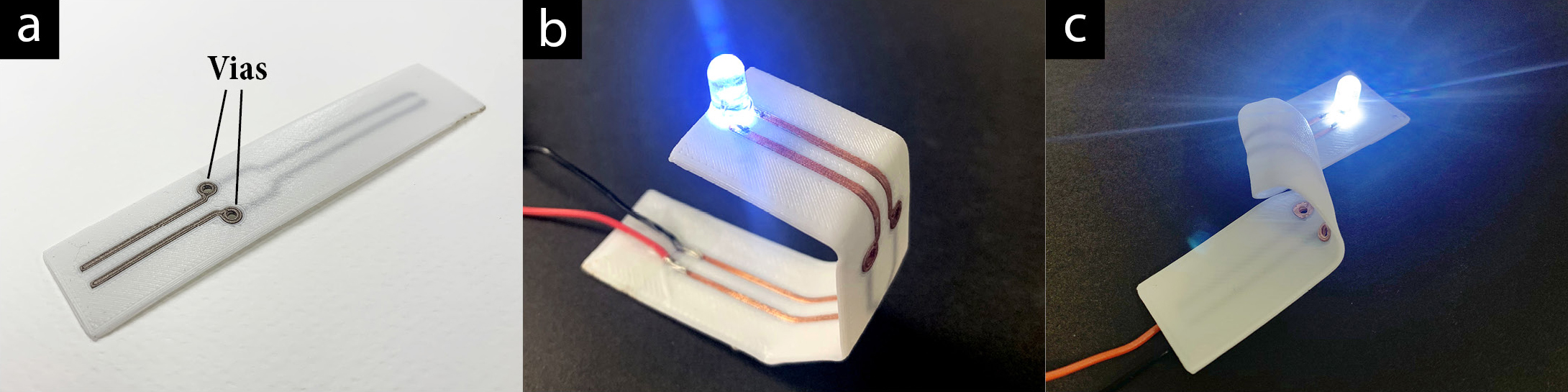}
  \caption{A basic TCB example: a) 3D printed PLA substrate and conductive traces with vias; b) heat bent double-sided LED circuit with electroplated circuit; c) heat twisted LED circuit with electroplated circuit}
  \label{fig:LED}
  \Description{This figure shows the example of TCB LED circuit board. It shows from the left to right: a) Flat 'as printed' TCB board, a heat bent electroplated LED circuit with two vias, heat twisted electroplated LED circuit with two vias}
\end{figure}


\subsection*{Impact of Thermoforming on the Conductivity of Trace}
\begin{figure}[h]
  \centering
  \includegraphics[width=1\linewidth]{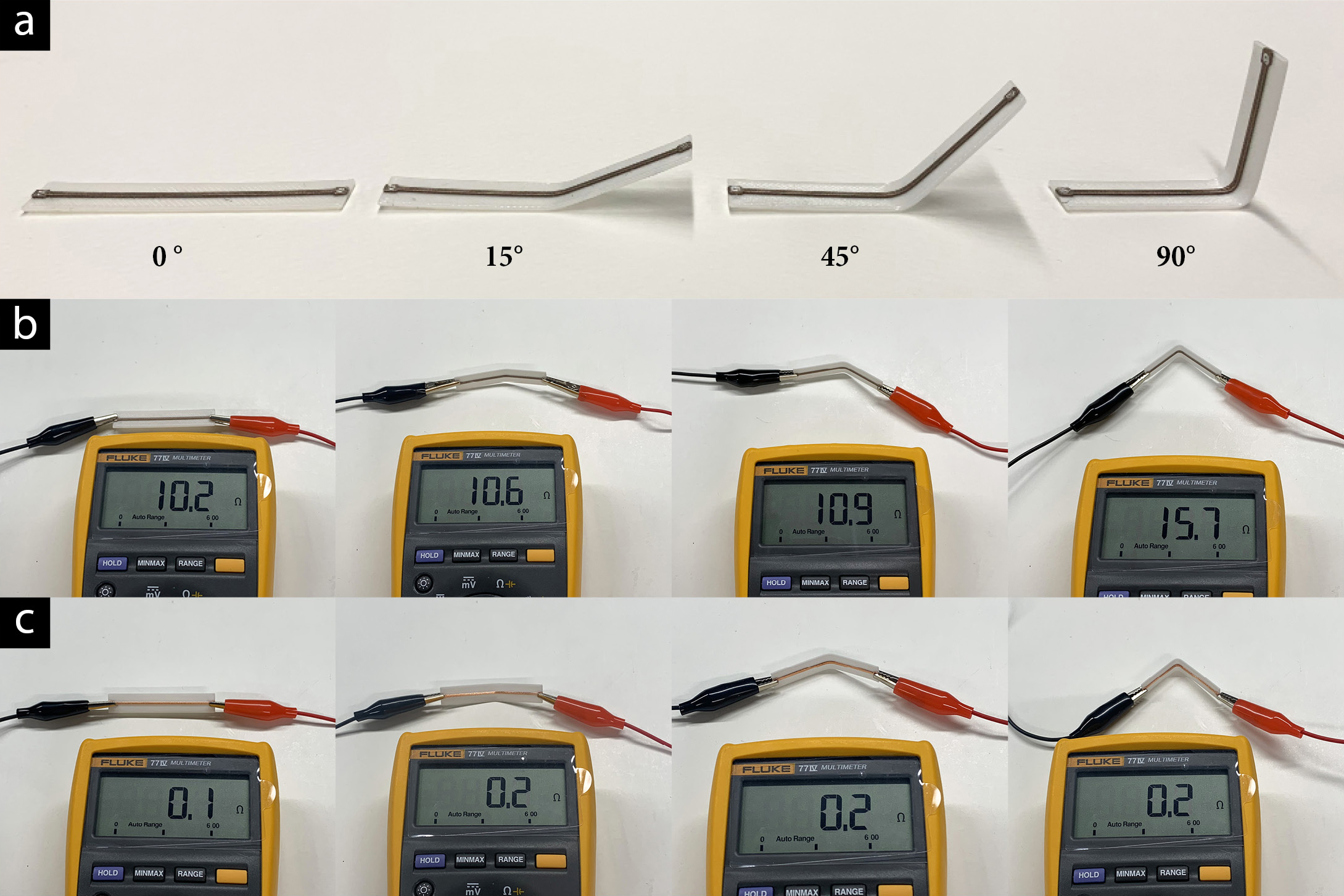}
 \caption{Resistance of the printed trace after each fabrication steps measured by digital multimeter. From left to right, the bending angles are: $0\degree$, $15\degree$, $45\degree$~and $90\degree$. a) Printed traces with varying bending angles; b) resistance of the traces after bending; from left to right: $10.2\Omega$, $10.6\Omega$, $10.9\Omega$, $15.7\Omega$; c) resistance of the traces after copper electroplating; from left to right: $0.1\Omega, 0.2\Omega, 0.2\Omega, 0.2\Omega$}
  \label{fig:ResistanceMeasuring}
  \Description{a) Pictures of a thin conductive PLA samples being heat bent into various angles. B) resistance of the heat bent samples shown on by the multimeter  c) resistance of copper electroplated samples shown by the multimeter}
\end{figure}

To study the changes occurring in the conductivity of the printed trace after thermoforming and electroplating, we printed 4 samples with 50~mm long conductive traces (1.3~mm wide and 0.5~mm tall). We first thermoformed the sample at varying angles: 0\degree, 15\degree, 45\degree~and 90\degree. Next, we copper electroplated the samples in a plating bath for 30 minutes at 0.4~V constant voltage (current range between 0.15 \& 0.22~A). To measure the resistance of the trace prior to electroplating, we applied drop of conductive ink to each end of the sample, as per the guideline provided by the manufacturer of the conductive filament~\cite{Multi3D}. As shown in Figure \ref{fig:ResistanceMeasuring}, the resistance of the sample increased as the bending angle increased. After copper electroplating the trace using the methods by Lazarus et al.~\cite{Lazarus2019}, the resistance of the samples reduced to 0.1~$\Omega$ for 0\degree~ and 0.2~$\Omega$ for the rest. The measured conductivity of the flat trace before and after electroplating was $7.54\times10^3$ S/m and $7.69\times10^5$ S/m, respectively. The electroplated conductive trace surface temperature increases to 30\degree C when 5A current flows 
~\cite{Kim2019}.
We expect that the resistance of the printed trace can be further reduced by lengthening the plating period. The examples shown in Figure~\ref{fig:LED} indicate that it is possible to copper electroplate the thermoformed trace even after extensive bending.

\section{TCB Resolution}

\begin{figure}[h]
  \centering
  \includegraphics[width=1\linewidth]{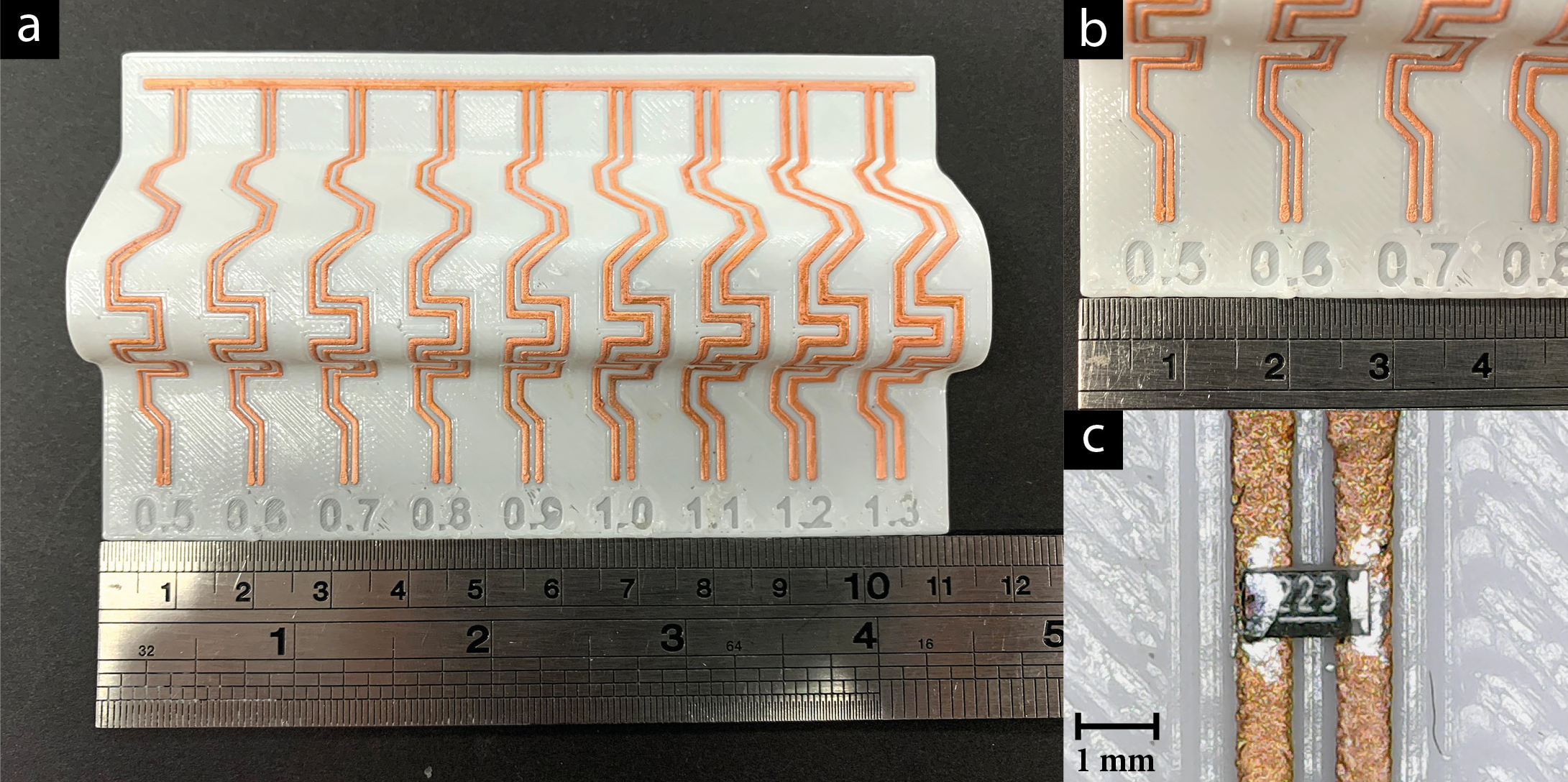}
  \caption{a) Top down view of the sample chart, b) Zoomed in view of the chart, c) 0603 SMD resistor on TCB trace with 0.5 mm spacing }
  \label{fig:smd}
  \Description{a) this figure shows various printed and electroplated traces with varying size. b) This image shows zoomed in view of the chart with a ruler to compare. c) shows microscopic image of SMD resistance mounted on thin copper electroplated traces with 0.5mm spacing.}
\end{figure}

While designing TCBs circuits, we were curious to find the smallest possible 3D printable width of the trace and spacing using conductive PLA, and whether it can still be thermoformed and electroplated. We printed a sample chart with varying trace widths and spacing. We printed a single layer of thin traces ranging from 0.5~mm in width and spacing up to 1.3~mm in width and spacing. Each time the width and spacing increased 0.1~mm, and all trace lengths were 93~mm. We chose to start from 0.5~mm width because it is arguably the smallest possible width a 0.4~mm nozzle can extrude at constant thickness. 0.4~mm is also the most common nozzle dimension found in FDM printers. For plating, we immersed the sample into the electrolyte solution for 120 minutes with a constant 0.4~V supply. As shown in \ref{fig:smd}, we were able to electroplate the 0.5~mm wide thermoformed conductive trace. 

As shown in Figure~\ref{fig:smd}c, a 0603 package type SMD resistor (1.5~mm x 0.8~mm) can fit between the trace with 0.5~mm spacing. This indicates that TCBs can accommodate small components and thus is useful to construct compact circuits.


\section{TCB Fabrication Workflow}
\begin{figure*}[!h]
  \centering
  \includegraphics[width=1.0\linewidth]{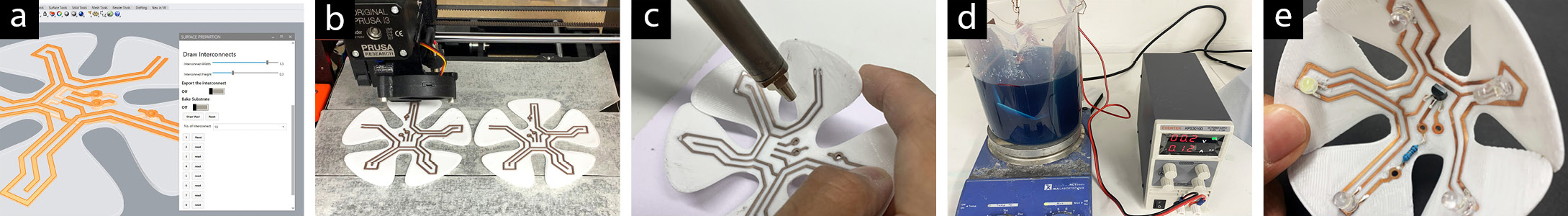}
  \caption{Example of TCB fabrication process: a) Designing the substrate and the traces using TCB parametric CAD tools; b) 3D printing the substrate and the traces using PLA and conductive filament; c) Thermoforming printed part using hot-air blower; d) copper electroplating in electrolyte solution bath; e) placement of electrical components}
  \label{fig:workflow}
  \Description{overall workflow of the TCB fabrication shown as a pictures a) screen capture of the CAD environment, b) printing process using prusai3 FDM printer, c) bending the plastic using hot-air blower, d) printed part immersed inside the copper electroplating solution bath, e) copper plated circuit with electrical components (resistor, transistor and white 5 mm LEDs placed }
\end{figure*}

The fabrication process of the TCB occurs in 5 stages: i) designing the PCB in a CAD environment with our plugin, ii) 3D printing the circuit and the substrates using an FDM printer, iii) thermoforming the printed parts with a hot-air blower, iv) copper electroplating the printed traces, and v) placing the electrical components using conductive ink and glue (Figure \ref{fig:workflow}).  
\subsection*{Stage 1: Designing TCB Editor}
Our design tool is built within the 3D modelling software with visual scripting extensions, namely \textit{Rhino3D} and \textit{Grasshopper}. We use the parametric functions of the \textit{Grasshopper} to design and edit the circuit layout optimized for 3D printing. First, the designer must import the 3D model of the substrate into the CAD environment. Next, with the graphical interface, the designer can draw the circuit elements directly onto the imported 3D model of the substrate. During this process, the designer can define the width and height of the trace as well as the position and size of the vias and sockets. Once completed, the software can export the substrate and the traces as \textit{stl} files for 3D printing.
\subsection*{Stage 2: 3D printing using conductive filament}
In this work, we used a Prusa i3 MK3S 3D printer (Prusa Research)~\cite{PRUSAPRINTER} with white PLA and conductive filament. Since TCB printing process only requires switching the filaments 4 times (max), instead of using dual-material 3D printer, we added a virtual extruder on the \textit{PrusaSlicer} and improvised the tool, i.e. changing G-code to manually switch between PLA and conductive filament. The benefit of using a single extruder instead of a dual extruder is that it can avoid stringing and oozing between printing path which can cause cross contamination of materials resulting in short circuits. To swap the materials, we used Prusa specific G-codes (\textit{M104} function), which changes the temperature of the extruder to the next material, and the \textit{M600} function to pause the printer to unload and feed in the new filaments, respectively. 
One issue we encountered during the switching process is that the copper-based conductive PLA leaves larger amounts of residue inside the extruder when unloading the filament. Therefore, we encourage the prospective user to be attentive when switching conductive filament to PLA by visually assessing that the copper based residue is fully cleared out of the nozzle and clean white PLA is being extruded before resuming the print. 
\textit{Printing parameters:} We tested printing the trace with 0.4~mm nozzle at 0.25~mm to 0.3~mm layer height and 0.8~mm to 1.3~mm width. In all cases the conductive trace with single layer height was conductive enough to be electroplated. For extruding copper based conductive filament, we set the temperature of the extruder to 150~\degree C with printing speed at 10~mm/s. For extruding the PLA, we set the temperature of the extruder at 205~\degree C and the printing speed at 45~mm/s. The temperature of the printing bed was set constant at 55 \degree C.
\subsection*{Stage 3: Heat Bending}
For heat bending the printed parts, we use a hot-air blower at 160~\degree C. Due to the low glass transition temperature of the PLA and the thinness of the parts, both PLA and the conductive filament became rubbery within a minute of blowing the hot air. Once it has become rubbery, we can shape the printed part and let it cool down until it became rigid again (Figure \ref{fig:workflow}c). Alternatively, we were able to therefrom the parts using a hairdryer, but this was not as precise in targeting small areas.
\subsection*{Stage 4: Copper Electroplating}
\begin{figure}[h]
  \centering
  \includegraphics[width=1\linewidth]{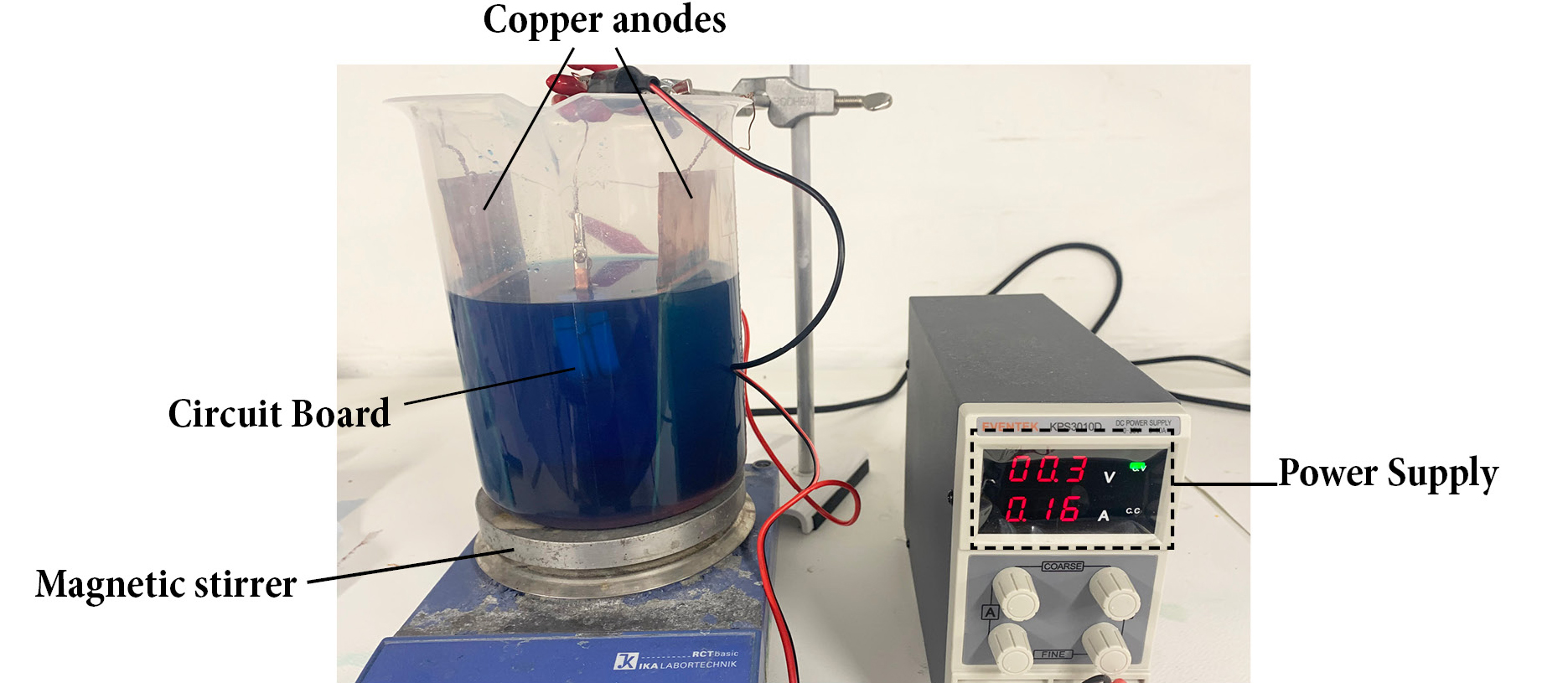}
  \caption{Copper electroplating bath setup with two copper anodes, immersed circuit board, DC power supply and magnetic stirrer}
  \label{fig:Plating}
  \Description{Image shows a copper electroplating solution filled inside a beaker. The bath is sitting on the magnetic stirrer. Two copper anode is connected to the power supply, the sample is immersed and visible translucently in the picture. Power supply shows the reading 0.03V and 0.16 A}
\end{figure}

Once thermofomred, the part is now ready to be immersed into copper plating bath. The plating solution is made up of copper sulphate, sulphuric acid, water and additive brightener. Following the general guide of copper electroplating, we used constant voltage between 0.2 to 0.4~V with varying current range according to the size of the circuit. We began electroplating at constant low voltage ($\sim$0.2V) until we observed that copper formed across the length of the circuit ($\sim$20 minutes). As more copper formed, we increased the current flow at constant voltage, increasing to 0.3V then to 0.4V ($\sim$40 minutes). This reduces the amount of current otherwise forced through the PLA, which may lead to joule heating and deformation. Comprehensive analysis of copper deposition rate on different conductive PLA can be found in Kim et al.'s paper~\cite{Kim2019}. Throughout plating, the bath was agitated with a magnetic stirrer at 400 rpm (Figure \ref{fig:Plating}). We used crocodile clips and copper wires to form continuous electrical contact before immersing the part into the bath. For small samples, we found that plating for 60 minutes formed adequate copper onto the printed traces.
\subsection*{Stage 5: Assembling}
\begin{figure}[h]
  \centering
  \includegraphics[width=1\linewidth]{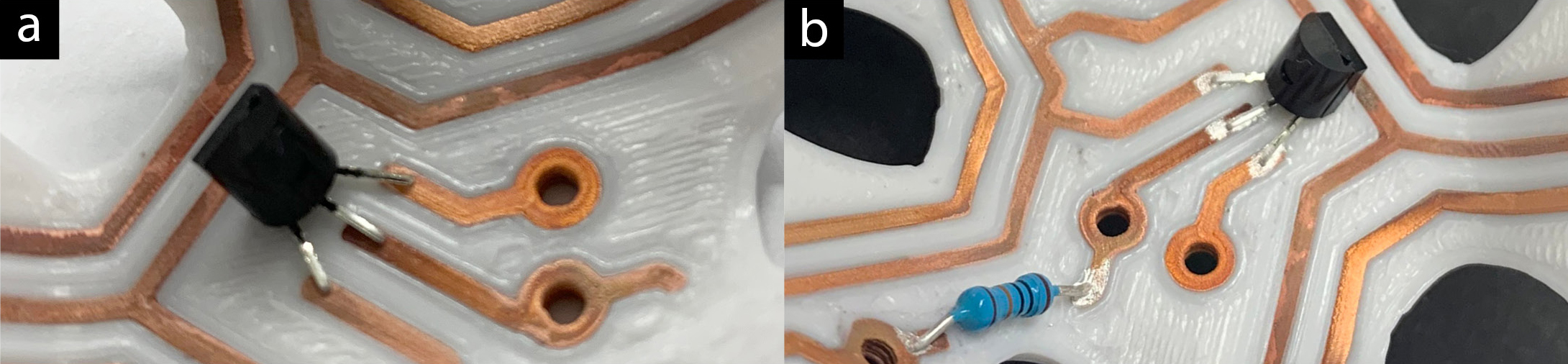}
  \caption{a) Transistor is first placed on the traces, b) conductive silver paste is then applied onto the leads}
  \label{fig:Placement}
  \Description{Zoomed in picture of the circuit board and the component a) transistor sitting on the plated traces but without any adhesive, b) shows transistor and resistor adhered to the plated circuit board with silver paste}
\end{figure}

\begin{figure}[h]
  \centering
  \includegraphics[width=0.5\linewidth]{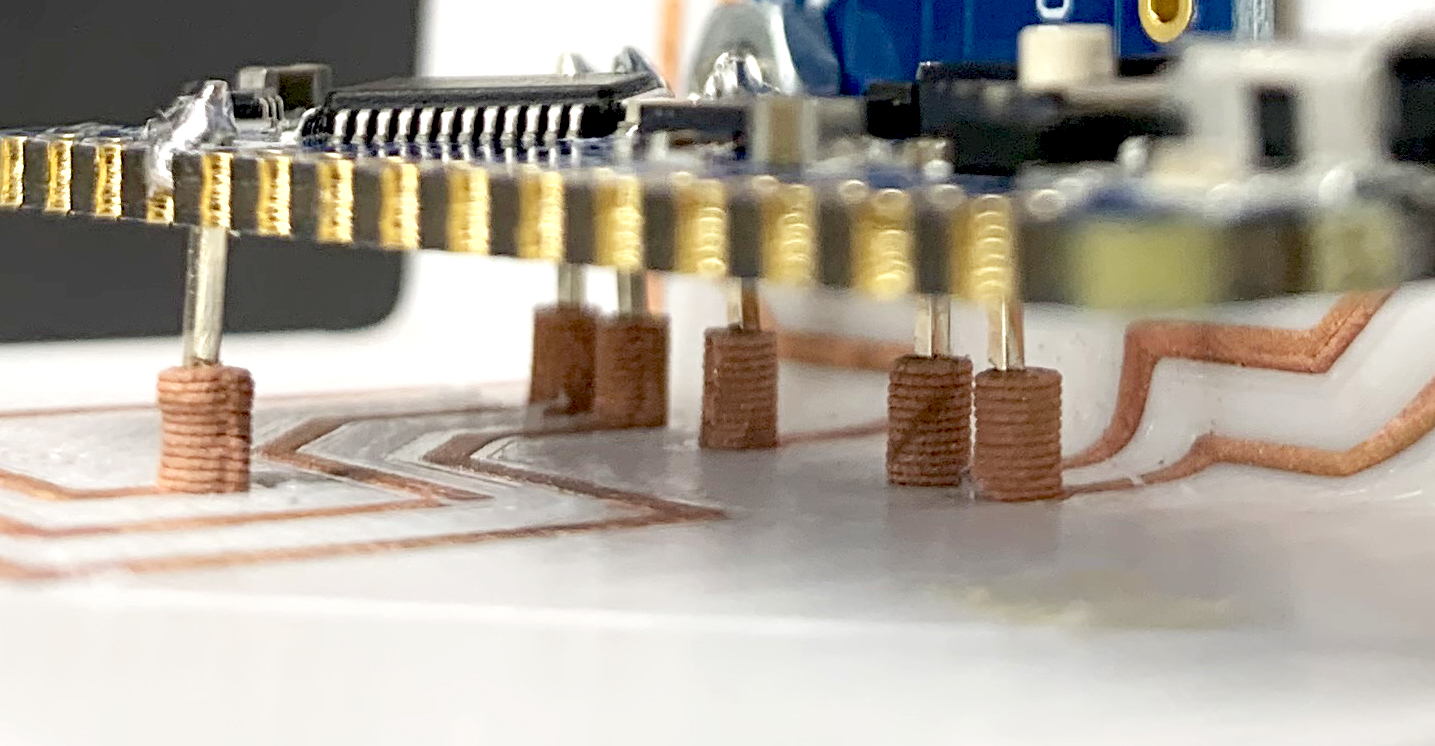}
  \caption{Arduino nano's header pins are inserted into the printed sockets}
  \label{fig:Sockets}
  \Description{The image shows ArduinoNano with pin headers being inserted into 3D printed and electropalted sockets}
\end{figure}

For assembly, we first position the electrical component on TCB. We then apply conductive silver paste with a syringe to adhere the leads of the components on the trace. 
Conductive silver ink reduces the contact resistance between the leads of the electrical component and the sockets. The differences in measured resistance at the socket before and after applying the silver ink was 0.1~$\Omega$. Once the conductive paste is dried and cured, we add a layer of superglue to fix them in position (Figure \ref{fig:Placement}).
There are many types of inexpensive commercial conductive silver paste that are curable at room temperature, often used for fixing broken PCB traces. In this paper, we used L100 conductive silver paste (Kemo-electronic) which has approximate resistance of 0.02~to $0.1~\Omega\cdot$cm$^{2}$. The conductive silver paste was dried at room temperature for 20 minutes before being covered with layered superglue on top. 
For attaching a microcontroller, we used cylindrical sockets with 1~mm radius and 2 to 3~mm height. We then press-in the header pins of the microcontroller into the sockets (Figure \ref{fig:Sockets}). To ease the pressing process, we sharpened the tips of the header pins with a cutter.
It is also possible to solder the components onto the electroplated traces, however, due to the low melting point of the thermoplastic, it is challenging to solder yielding a consistent finish.
\section{Response to mechanical deflection}
While testing the high current application, we accidentally damaged the electroplated trace by deflecting the substrate too severely. As the plated layer fractured we noticed an increase in temperature, which led to burning the trace and substrate. To investigate the assumed cause and identify limits, we performed a 3-point bending test on three printed samples to find the flexural strain on the substrate at which the trace would fracture. Each sample consists of PLA substrate (70 mm long, 10 mm wide and 0.9 mm deep (\textit{d})) and single layer of conductive trace in the substrate (70~mm long and 1.0 mm wide). The samples were copper electroplated for 60 minutes with constant 0.4~V. The sample was positioned on two supporting pins that are 40 mm (\textit{L}) apart. To achieve precise displacement of the loading pin, we used a 3-axis positioning stage controlled by a Mach3 machine controller. We incrementally increased displacement by 0.25~mm (\textit{D}) each step. 
\begin{figure}[h]
  \centering
  \includegraphics[width=1\linewidth]{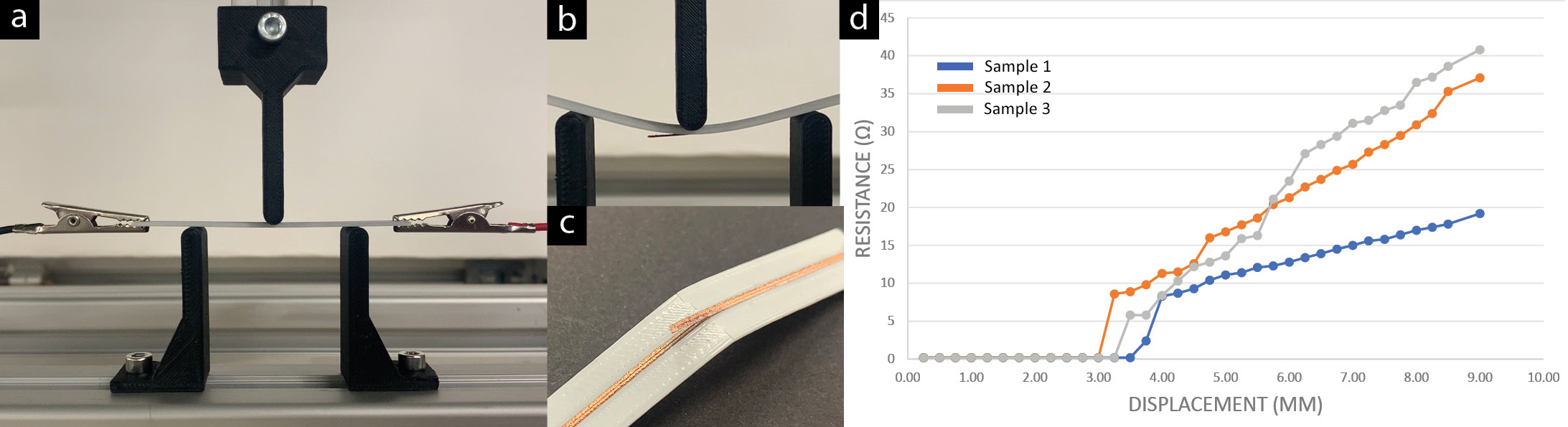}
  \caption{a) Flexural test setup; b) Deflection of the center of beam when electroplated layer fractured; c) close up view of fractured plating; d) Electroplated trace flexural test result }
  \label{fig:flex}
  \Description{a) shows overall setup of the flexural test. Two black supporting pin and one pressing pin is holding the sample flat. The sample is also connected to two alligator clips, b) a zoomed in picture of the supporting pin pressing the sample. This image also shows the de-laminated plated layer. c) shows zoomed in picture of the fractured copper plated trace d) shows a graph plotted for  resistance against the displacement. It show where the fractures are happening three curves representing three samples are shown.}
\end{figure}
As shown in Figure~\ref{fig:flex}d, resistance of the trace remained the same until the first fracture of the plating, identifiable by the sudden increase in resistance. After the initial fracture, resistance rose quickly towards the point at which the conductivity of the sample was reliant on the unplated conductive PLA. 
For Sample 1, when the fracture occurred the flexural strain ($\varepsilon_f=~\frac{6Dd}{L^2}$) was 0.0127, displacement was 3.75 mm and the obtuse angle between the pins was 158.76$\degree$. For Sample 2, flexural strain was 0.0110, displacement was 3.25 mm and the obtuse angle between the pins was 161.54$\degree$. For sample 3, flexural strain was 0.118, displacement was 3.5 and angle the obtuse angle between the pins was 160.15$\degree$. We conclude without considering thickness that bending the electroplated part in excess of 18.46$\degree$ parallel to the orientation of the trace will begin to cause damage to the plating. The orientation of the trace, therefore, should be considered around the parts of a device that are prone to movements and deflections. Adopting a stretchable pattern could also be considered.

\section{TCB Design Editor}
\subsection*{Pitch dimension}
\begin{figure}[h]
  \centering
  \includegraphics[width=1\linewidth]{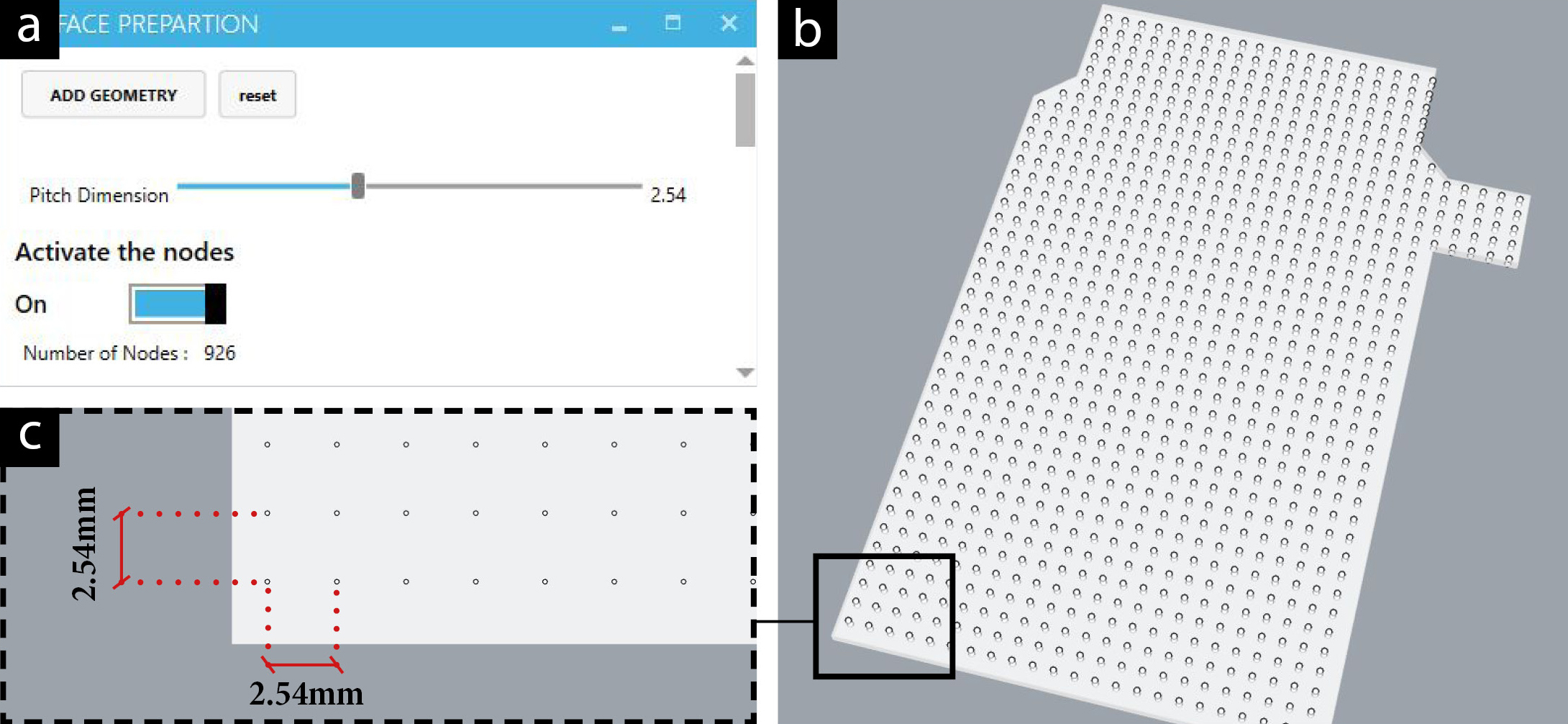}
  \caption{a) TCB design editor interface; b) imported 3D model of the substrate with point-grid; c) Close up view of the point grid with pitch dimension}
  \label{fig:PitchEditor}
  \Description{A 3D modelling environment is shown in the images a) graphical user interface for adjusting the pitch dimension, b) Overall view of the surface with generated pitch, c) close up view of the points with the distance (2.54mm) indicated as lines}
\end{figure}

Designing TCBs begins by importing the 3D model of the substrate into the editing interface in \textit{Grasshopper}. Once imported, \textit{Grasshopper} automatically generates a point-grid on the top and bottom layers of the substrate. These points are used to draw the circuit layout. Before drawing the circuit, the pitch dimension must be defined (Figure~\ref{fig:PitchEditor}). Pitch dimension is the distance between each point in the grid, and ultimately determines how finely the traces can through be drawn. More complex circuitry may require smaller pitch dimension and vice versa (e.g. the common pitch for headers and pins on through-hole package microcontrollers and breadboards is 2.54~mm).
\subsection*{Traces and Vias}
 Once the pitch is set, the designer can draw conductive traces by clicking the number of interconnects and selecting a set of points on the surface. After selection, the 3D trace is automatically created and trims the substrate accordingly. More traces can be added whilst paramtrically adjusting the width and height of the conductive traces at any point during the process using a slider. To connect the conductive traces between top and bottom layers, a designer can click the \textit{Add Vias} button and select the points at which to locate the vias. Like the previous step, the size of the vias can be parametrically adjusted. Sockets may be added where header pins or IC leads will be placed. Sockets are added by simply clicking the points and the size can be adjusted using the slider.

\begin{figure}[h]
  \centering
  \includegraphics[width=1\linewidth]{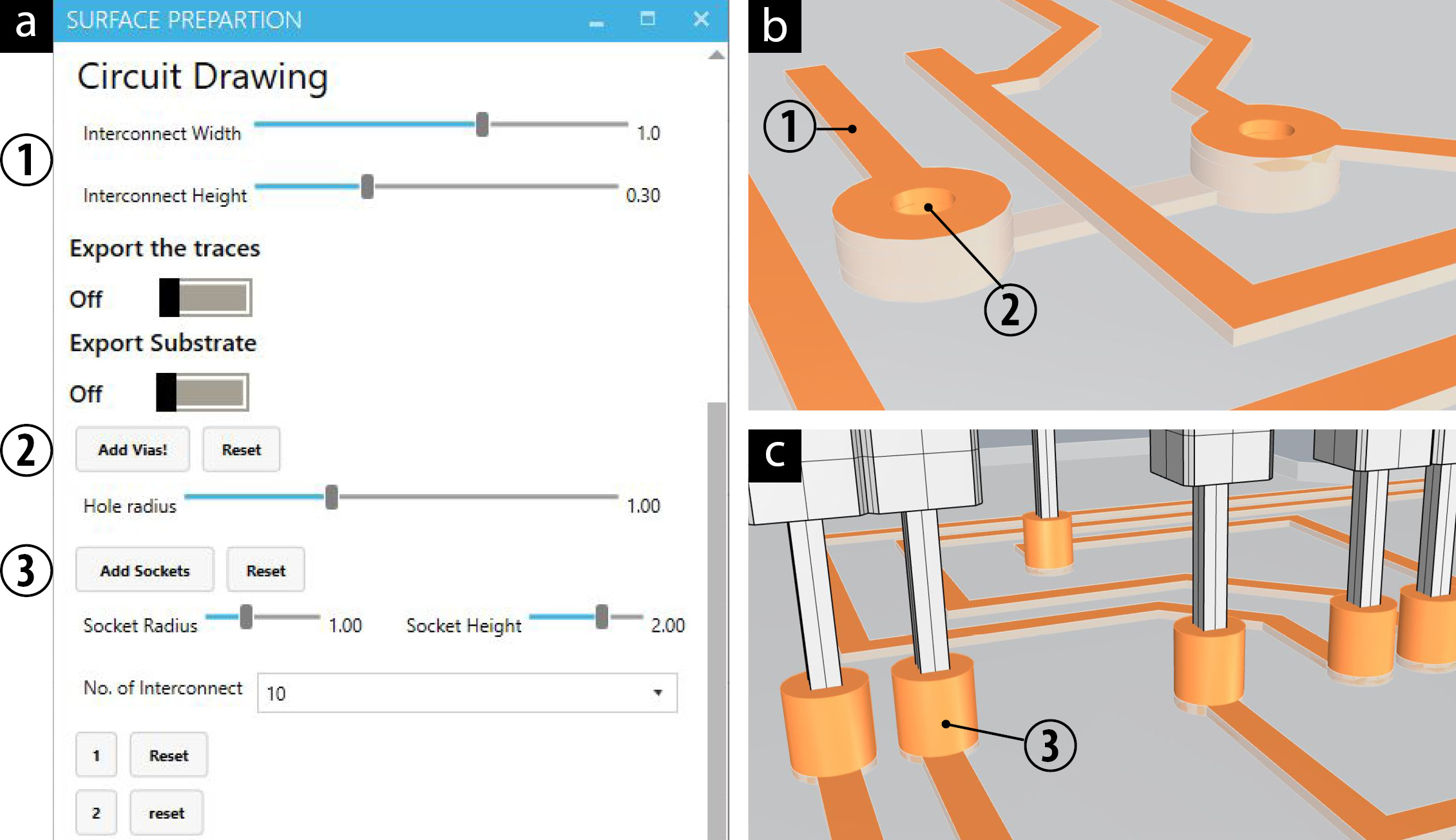}
  \caption{a) TCB design editor interface; b) traces and vias; c) sockets}
  \Description{3D modeling envrionement is shown a) graphical user interface window showing buttons and sliders for drawing circuits 1) trace 2) vias and 3) sockets, b) shows the perspective image of constructed traces and vias c) shows perspective image of sockets and header pins}
\end{figure}
\subsection*{Exporting printable designs} 
 Once the design is complete, a designer can export the substrate and traces by clicking the \textit{Export} tab. The geometry can then be outputted into STL files using the Rhino export system for slicing.
 
\section{Applications of TCB}
In this section, we showcase various example devices made with TCB. These examples are chosen to demonstrate TCB’s unique capabilities in terms of form factors, electrical characteristics and interaction mechanisms. 

\subsection*{TCB works with Complex Geometry}
\begin{figure}[t]
  \centering
  \includegraphics[width=1\linewidth]{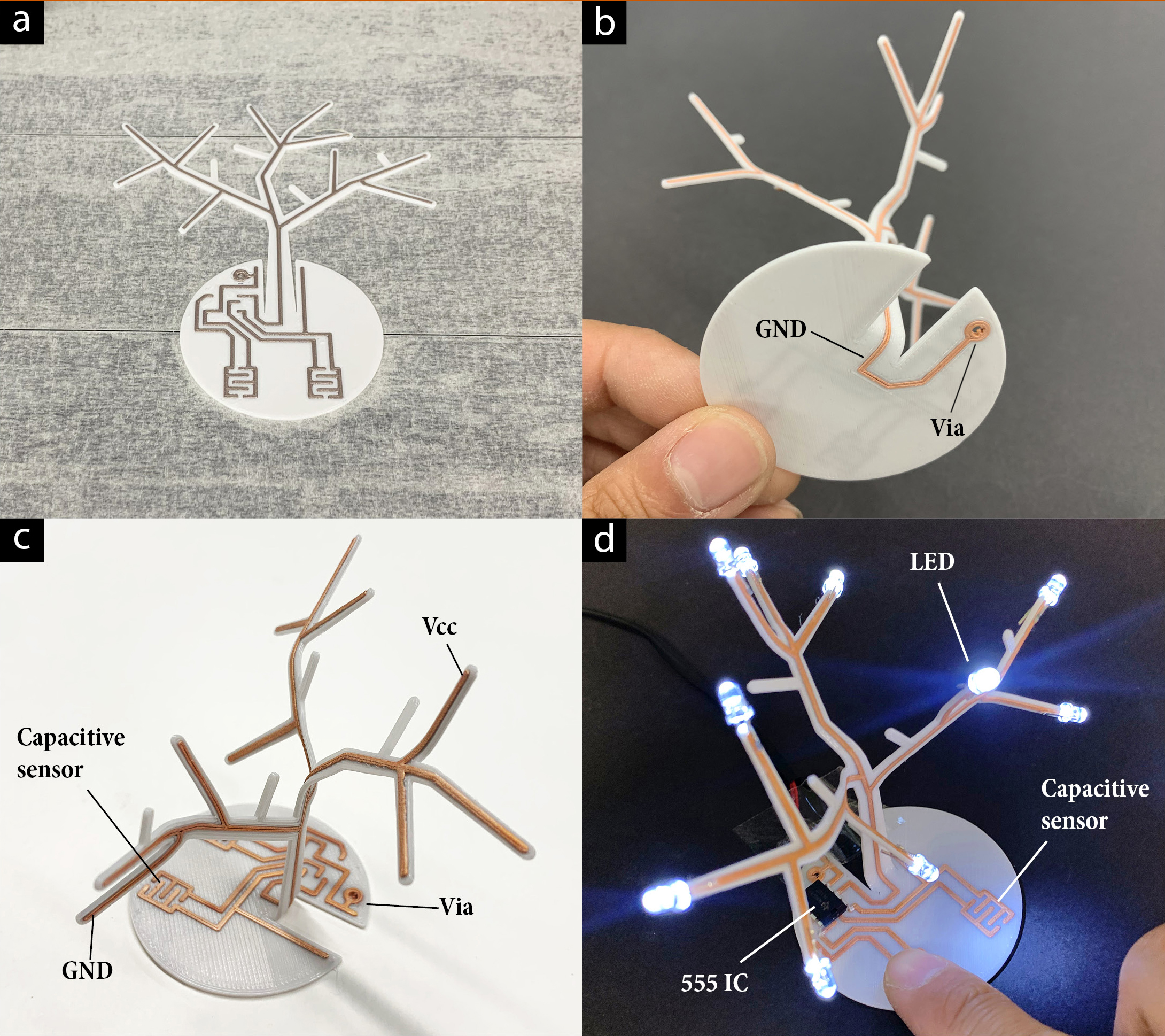}
  \caption{a) Tree lamp as printed on printing bed; b) bottom layer of the tree lamp and GND trace; c) tree lamp has 3D branch structure that is heat bent in various direction and angles, which would be time consuming and technically difficult for conventional fabrication; d) Fully assembled tree lamp}
  \label{fig:tree}
  \Description{This figure shows the making process and the operation of the tree lamp a)as printed view of the tree lamp circuit, b) bottom layer of heat bent and electroplated circuit c) top layer of heat bent and electroplated circuit d) a bright tree lamp in switched on with a finger tab}
\end{figure}

\textit{Tree Lamp} (Figure~\ref{fig:tree}): Constructing rigid and thin branch structures in 3D is a challenging task for any form of fabrication. Here, we showcase a double-sided, tree-like 3D circuit board that is carrying both GND and Vcc traces along confined branches. Since the insulating layer and printed traces are thermoformed together as a single entity, regardless to the amount of bending, the circuit is protected from potential shorts. 3~mm LEDs are simply clipped onto the branches with their anode and cathode leads pressing against the Vcc and GND traces at either side of the branch (separated by the PLA in the middle). \textit{Tree lamp} can be switched on and off using the capacitive sensors controlled by 555 timer IC. 
Constructing freeform circuit boards like this tree lamp, which has lots of undercutting and overlapping form factor, has been thus-far impossible using traditional subtractive and additive manufacturing techniques. Thus, we identify the tree lamp as a unique form factor that is \emph{only} manufacturable using our TCB technique.
\subsection*{TCB Keeps Things Seperate}
\begin{figure}[h]
  \centering
  \includegraphics[width=1\linewidth]{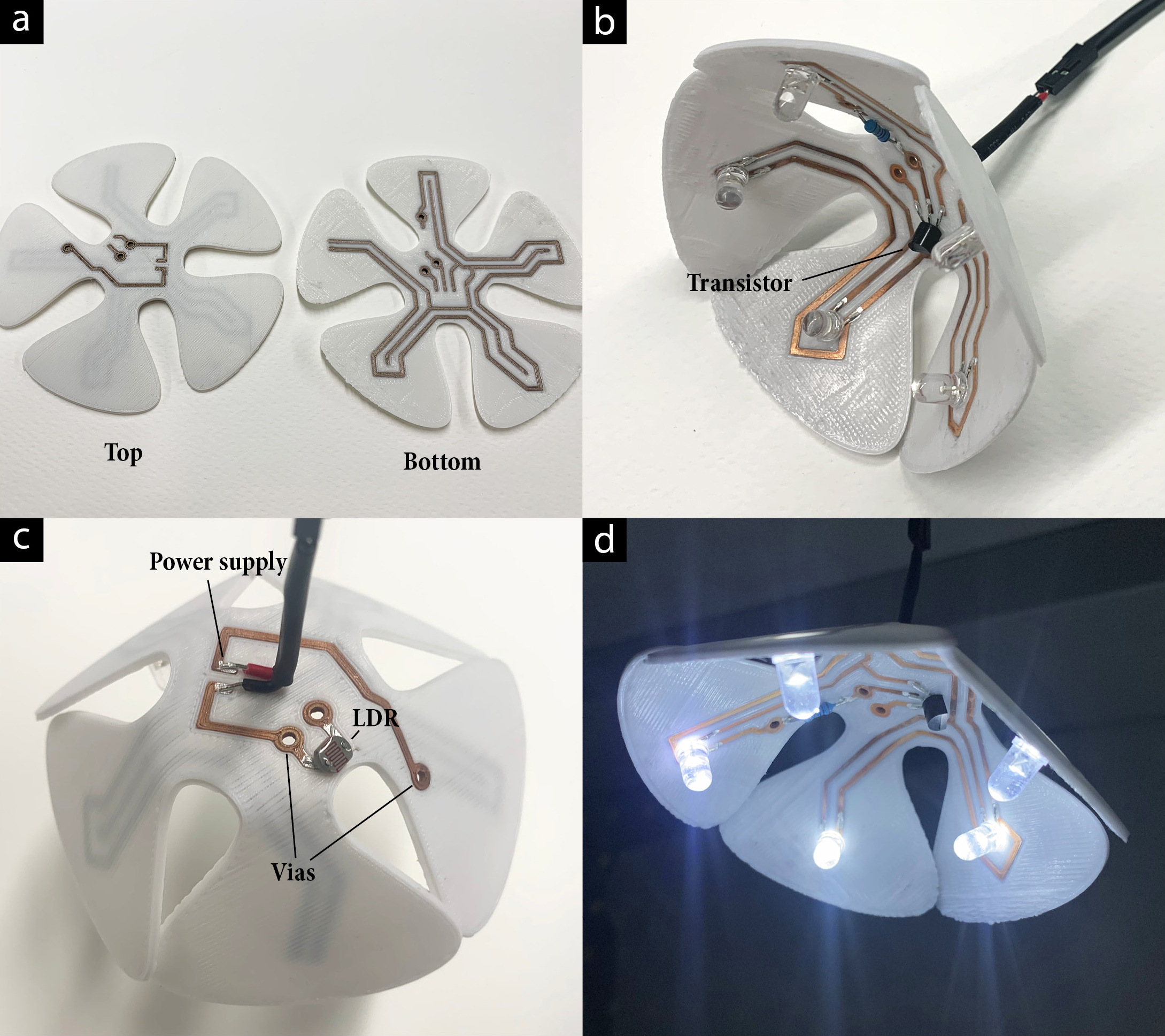}
  \caption{a) Pendant light as printed; b) bottom layer as an actuating layer; c) top layer as an input layer with power supply and LDR sensor; d) Pendent Light in action}
  \label{fig:pendent}
  \Description{construction process and the operation of the light dependent pendent light is shown a) flat as printed view of the circuit, b) bottom layer of thermoformed, electroplated and assembled pendent light, c) top layer of pendent light d) bright pendent light in dark environment }
\end{figure}

\textit{Pendant Light} (Figure~\ref{fig:pendent}): One benefit of using double sided circuit boards is the greater freedom of component organisation. Here we used double-sided capability of TCB to dedicate all the input elements, including power supply and light dependent resistor (LDR), to the top layer, and the actuating elements to bottom layer. This is a light sensitive pendant lamp which turns on and off using an LDR and a transistor as a switch. If the LDR sensor and LEDs were positioned on the same surface, the lamp would malfunction through self interference. There are many applications where sensing and actuating elements, for example, may benefit from being kept separated. Having greater freedom in terms of components placement thus provides designers with more flexibility when choosing form factors of the 3D interactive objects.
\subsection*{TCB Handles High-Current Applications}
\begin{figure}[h]
  \centering
  \includegraphics[width=1\linewidth]{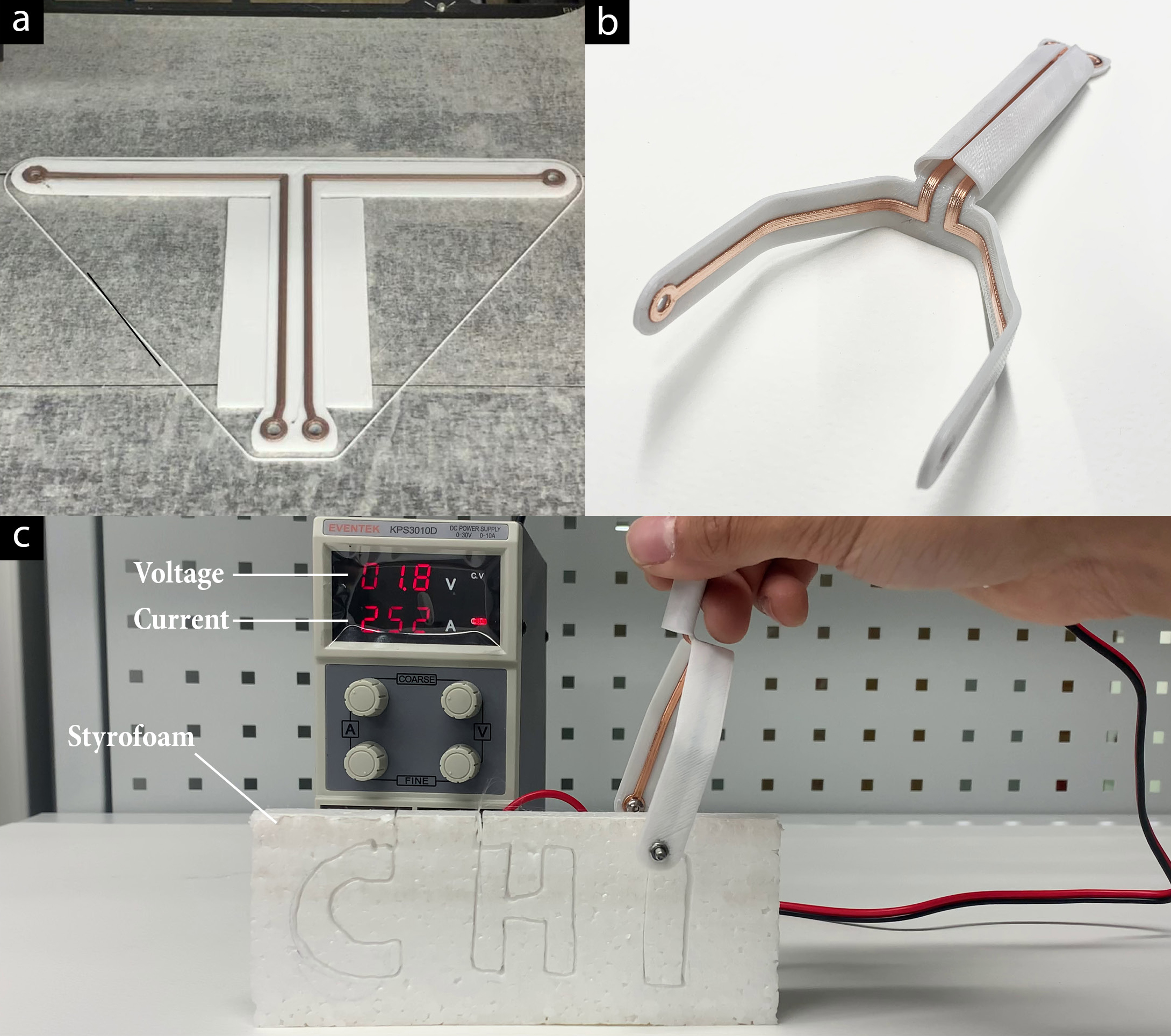}
  \caption{a) Hot-wire form cutter as printed; b) thermoformed and electroplated circuit board; c) 3D printed hot-wire foam cutter in operation, the screen on DC power supply indicates 2.52~A of constant current is being carried by the printed traces }
  \label{fig:Hot}
  \Description{construction process and operation of the TCB hot-wire cutter is shown a) as printed circuit b) thermoformed and electroplated c) assembled and operating image of the hot wire foam cutter, cutting the shape 'CHI' out of styrofoam block with 0.18 V and 2.52 A }
\end{figure}

\textit{Hot-wire Cutter} (Figure~\ref{fig:Hot}): In the HCI community, examples of using conductive materials for 3D interactive objects have been focused on low-current applications such as sensing and displaying. This is partly due to the low current carrying capacity of digitally fabricated materials, as well as the thermal properties of the substrates. By printing a thick layer of conductive trace and copper plating, we demonstrate unprecedented use of conductive filament trace for heating up 24 AWG Nichrome wire to 250\degree C to sculpt styrofoam. As shown in Figure\ref{fig:Hot}c, the hot-wire foam cutter is powered with a constant current of 2.52~A from the DC power supply. 

\subsection*{TCB for Rapid Prototypes}
\begin{figure}[t]
  \centering
  \includegraphics[width=\linewidth]{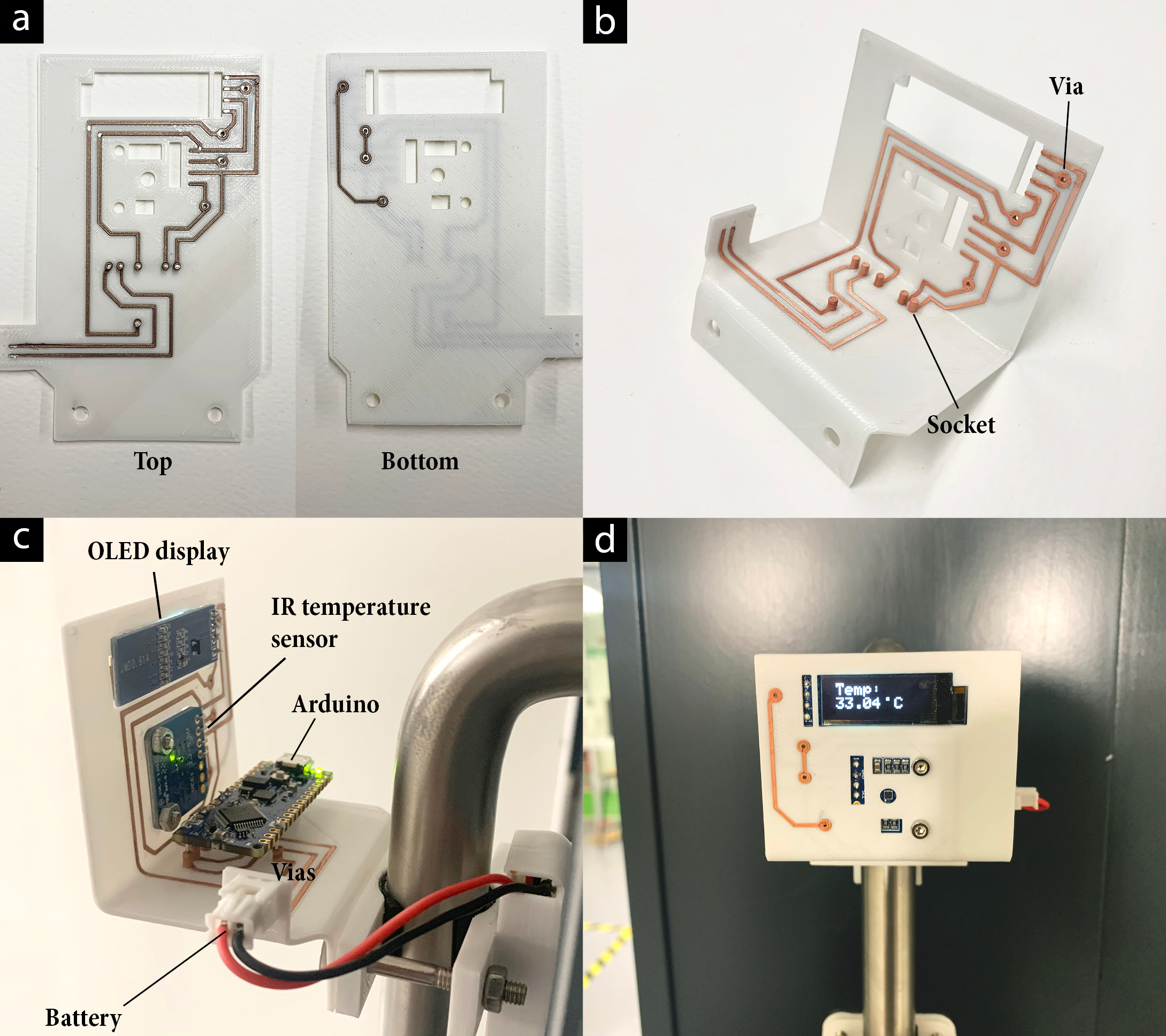}
  \caption{a) Contactless thermometer circuit board as printed; b) thermoformed and electroplated circuit board; c) assembled TCB with microcontroller and battery fixed to door handle; d) IR thermometer in action}
  \label{fig:Temperaturedetector}
  \Description{This figure shows the making process and operation of the contactless thermometer, a) as printed view of the circuit, b) thermoformed and electroplated circuit board, c) fully assembled circuit board showing arduino nano, portable battery connector, OLED Display and IR temperature sensor, d) shows the front view of the thermometer attached to a door handle}
\end{figure}

\textit{Contactless Infra-red Thermometer} (Figure~\ref{fig:Temperaturedetector}): 3D printers have demonstrated important utility for manufacturing functional objects during the COVID-19 pandemic. This is evidenced by rapid prototyping of face shields, masks and hands-free adaptors. With emerging functional materials like conductive filament, 3D printed end-use products can also be electrically interactive. To demonstrate this potential, we showcase rapid prototyping of freeform circuit boards that can be attached to a door handle whilst also containing electrical modules including a display, sensors and microcontroller. Here, the TCB output resembles the aesthetics of a MID device, but is manufactured using a comparably minimal process, and in an economical way. This contactless thermometer is controlled by an Arduino nano over I2C and powered with a 3.7~V battery. 
\section{Limitations and Future Work}

 \textbf{Flexural Strength.} As evaluated in the Section 6, the copper plated trace can be fractured and de-laminated under flexural strain. While operating the first sample for hot-wire foam cutter, we noticed that fractures occur due to the deflection of the substrate. We found that for high current applications, fractures can cause burning of the conductive PLA due to increased resistance. One method of overcoming this issue is by increasing the thickness of the substrate. This requires additional 3D printed layers, however, thus increasing printing time and detracting from one of the main merits of the proposed method. In future work, we intend to examine the use of stretchable wire patterning, as demonstrated in {ElectroDermis}~\cite{10.1145/3290605.3300862}, to improve the flextural strain of the electroplated trace. We will investigate constructing TCBs using other materials like flexible thermoplastic polyurethane and shape-changing polymers to expand the applicability of TCB. Furthermore, we will investigate and characterize the mechanical (and electrical) properties of TCBs against relevant industrial standards with a view to ensuring safety and reliability.

 \textbf{Electroplating Bath.} The scale of TCB fabrication under the current method is restricted by the size of the available plating bath. In this work, we used 1.4~L of electrolyte solution in a cylindrical beaker with two copper anodes and stirring apparatus, which provided immersible bounding volume of 90~mm x 120~mm x 100~mm. Although this volume was just enough for immersing the examples shown, this restricts plating parts for larger objects. To circumvent this issue, we will evaluate the feasibility of using brush plating.

 \textbf{Accurate and repeatable manufacturing.} The current method of thermoforming printed parts using a hot air blower allows the designer to intuitively bend the parts into 3D shapes. This method may, however, suffer from repeatability issues for more complex scenarios. We will explore further industrialized thermoforming techniques, such as vacuum forming and pressure forming, which can accurately and repetitively construct 3D shapes in molds. We will investigate deformation characteristics of thermoforming the 3D printed PLA sheet, and explore additional methods employing computational techniques to design and fabricate TCBs~\cite{Schuller2016ComputationalThermoforming}.

\section{Conclusion}
In this paper we presented Thermoformed Circuit Boards together with a novel approach to their construction yielding freeform, rigid and double-sided 3D circuit boards based on heat bending conductive PLA. We demonstrated that TCB is an inexpensive and highly accessible fabrication technique that enables rapid construction and exhibits good electrical performance. We demonstrated the applicability of TCB through a range of examples. These included designing TCBs for a variety physical form factors, electrical performance requirements, and interactive devices. We examined the electrical and mechanical properties of TCB devices, providing design insights for future TCB devices. We described and provided a new parametric design editor for TCBs, which allows designers to write circuit elements directly onto the substrate and make quick alterations through a graphical user interface. We showed that TCB can achieve fine trace resolution and space widths that can accommodate the use of SMD components, which we will explore in future work to construct more compact and complex TCB devices and objects. 
We hope TCBs can become a useful resource to the HCI community, broadening research participation in prototyping 3D printed electronics and artefacts. Although the technique extends known methods of prototyping electronically interactive objects, its relative simplicity, speed and cost effectiveness are attractive. We believe TCBs can be useful within various research areas including wearables, displays and robotics.

\bibliographystyle{ACM-Reference-Format}
\bibliography{references}


\begin{thebibliography}{43}


\ifx \showCODEN    \undefined \def \showCODEN     #1{\unskip}     \fi
\ifx \showDOI      \undefined \def \showDOI       #1{#1}\fi
\ifx \showISBNx    \undefined \def \showISBNx     #1{\unskip}     \fi
\ifx \showISBNxiii \undefined \def \showISBNxiii  #1{\unskip}     \fi
\ifx \showISSN     \undefined \def \showISSN      #1{\unskip}     \fi
\ifx \showLCCN     \undefined \def \showLCCN      #1{\unskip}     \fi
\ifx \shownote     \undefined \def \shownote      #1{#1}          \fi
\ifx \showarticletitle \undefined \def \showarticletitle #1{#1}   \fi
\ifx \showURL      \undefined \def \showURL       {\relax}        \fi
\providecommand\bibfield[2]{#2}
\providecommand\bibinfo[2]{#2}
\providecommand\natexlab[1]{#1}
\providecommand\showeprint[2][]{arXiv:#2}

\bibitem[\protect\citeauthoryear{??}{3DL}{[n.d.]}]%
        {3DLDS}
 \bibinfo{year}{[n.d.]}\natexlab{}.
\newblock \bibinfo{title}{{3D MIDs through laser direct structuring (LDS)}}.
\newblock
\newblock
\urldef\tempurl%
\url{https://www.lpkf.com/en/industries-technologies/electronics-manufacturing/3d-mids-with-laser-direct-structuring-lds}
\showURL{%
\tempurl}


\bibitem[\protect\citeauthoryear{??}{Mul}{[n.d.]}]%
        {Multi3D}
 \bibinfo{year}{[n.d.]}\natexlab{}.
\newblock \bibinfo{title}{{Multi3D}}.
\newblock
\newblock
\urldef\tempurl%
\url{https://www.multi3dllc.com/faqs/}
\showURL{%
\tempurl}


\bibitem[\protect\citeauthoryear{??}{Opt}{[n.d.]}]%
        {OptomecElectronics}
 \bibinfo{year}{[n.d.]}\natexlab{}.
\newblock \bibinfo{title}{{Optomec Aerosol Jet printing technology effectively
  produces 3D printed electronics}}.
\newblock
\newblock
\urldef\tempurl%
\url{https://optomec.com/printed-electronics/aerosol-jet- technology/}
\showURL{%
\tempurl}


\bibitem[\protect\citeauthoryear{??}{PRU}{[n.d.]}]%
        {PRUSAPRINTER}
 \bibinfo{year}{[n.d.]}\natexlab{}.
\newblock \bibinfo{title}{{PRUSA I3 MK3S 3D PRINTER}}.
\newblock
\newblock
\urldef\tempurl%
\url{https://www.prusa3d.com/original-prusa-i3-mk3/}
\showURL{%
\tempurl}


\bibitem[\protect\citeauthoryear{??}{Vox}{[n.d.]}]%
        {Voxel8Specifications}
 \bibinfo{year}{[n.d.]}\natexlab{}.
\newblock \bibinfo{title}{{Voxel8 Print specifications}}.
\newblock
\newblock
\urldef\tempurl%
\url{http://store.voxel8.com/faq}
\showURL{%
\tempurl}


\bibitem[\protect\citeauthoryear{Adams, Duoss, Malkowski, Motala, Ahn, Nuzzo,
  Bernhard, and Lewis}{Adams et~al\mbox{.}}{2011}]%
        {Adams2011}
\bibfield{author}{\bibinfo{person}{Jacob~J. Adams}, \bibinfo{person}{Eric~B.
  Duoss}, \bibinfo{person}{Thomas~F. Malkowski}, \bibinfo{person}{Michael~J.
  Motala}, \bibinfo{person}{Bok~Yeop Ahn}, \bibinfo{person}{Ralph~G. Nuzzo},
  \bibinfo{person}{Jennifer~T. Bernhard}, {and} \bibinfo{person}{Jennifer~A.
  Lewis}.} \bibinfo{year}{2011}\natexlab{}.
\newblock \showarticletitle{{Conformal printing of electrically small antennas
  on three-dimensional surfaces}}.
\newblock \bibinfo{journal}{\emph{Advanced Materials}} \bibinfo{volume}{23},
  \bibinfo{number}{11} (\bibinfo{year}{2011}), \bibinfo{pages}{1335--1340}.
\newblock
\showISSN{09359648}
\urldef\tempurl%
\url{https://doi.org/10.1002/adma.201003734}
\showDOI{\tempurl}


\bibitem[\protect\citeauthoryear{Angel, Tsang, Bedair, Smith, and
  Lazarus}{Angel et~al\mbox{.}}{2018}]%
        {Angel2018}
\bibfield{author}{\bibinfo{person}{Kristin Angel}, \bibinfo{person}{Harvey~H.
  Tsang}, \bibinfo{person}{Sarah~S. Bedair}, \bibinfo{person}{Gabriel~L.
  Smith}, {and} \bibinfo{person}{Nathan Lazarus}.}
  \bibinfo{year}{2018}\natexlab{}.
\newblock \showarticletitle{{Selective electroplating of 3D printed parts}}.
\newblock \bibinfo{journal}{\emph{Additive Manufacturing}}
  \bibinfo{volume}{20}, \bibinfo{number}{February} (\bibinfo{year}{2018}),
  \bibinfo{pages}{164--172}.
\newblock
\showISSN{22148604}
\urldef\tempurl%
\url{https://doi.org/10.1016/j.addma.2018.01.006}
\showDOI{\tempurl}


\bibitem[\protect\citeauthoryear{Burstyn, Fellion, Strohmeier, and
  Vertegaal}{Burstyn et~al\mbox{.}}{2015}]%
        {10.1007/978-3-319-22701-6_25}
\bibfield{author}{\bibinfo{person}{Jesse Burstyn}, \bibinfo{person}{Nicholas
  Fellion}, \bibinfo{person}{Paul Strohmeier}, {and} \bibinfo{person}{Roel
  Vertegaal}.} \bibinfo{year}{2015}\natexlab{}.
\newblock \showarticletitle{{PrintPut: Resistive and Capacitive Input Widgets
  for Interactive 3D Prints}}. In \bibinfo{booktitle}{\emph{Human-Computer
  Interaction -- INTERACT 2015}}, \bibfield{editor}{\bibinfo{person}{Julio
  Abascal}, \bibinfo{person}{Simone Barbosa}, \bibinfo{person}{Mirko Fetter},
  \bibinfo{person}{Tom Gross}, \bibinfo{person}{Philippe Palanque}, {and}
  \bibinfo{person}{Marco Winckler}} (Eds.). \bibinfo{publisher}{Springer
  International Publishing}, \bibinfo{address}{Cham},
  \bibinfo{pages}{332--339}.
\newblock
\showISBNx{978-3-319-22701-6}


\bibitem[\protect\citeauthoryear{Choong, Tan, Patel, Choong, Chen, Low, Tan,
  Patel, and Chua}{Choong et~al\mbox{.}}{2020}]%
        {Choong2020ThePandemic}
\bibfield{author}{\bibinfo{person}{Yu~Ying~Clarrisa Choong},
  \bibinfo{person}{Hong~Wei Tan}, \bibinfo{person}{Deven~C. Patel},
  \bibinfo{person}{Wan Ting~Natalie Choong}, \bibinfo{person}{Chun-Hsien Chen},
  \bibinfo{person}{Hong~Yee Low}, \bibinfo{person}{Ming~Jen Tan},
  \bibinfo{person}{Chandrakant~D. Patel}, {and} \bibinfo{person}{Chee~Kai
  Chua}.} \bibinfo{year}{2020}\natexlab{}.
\newblock \showarticletitle{{The global rise of 3D printing during the COVID-19
  pandemic}}.
\newblock \bibinfo{journal}{\emph{Nature Reviews Materials}}
  (\bibinfo{year}{2020}), \bibinfo{pages}{1--3}.
\newblock
\showISSN{2058-8437}
\urldef\tempurl%
\url{https://doi.org/10.1038/s41578-020-00234-3}
\showDOI{\tempurl}


\bibitem[\protect\citeauthoryear{Dichtl, Sippel, and Krohns}{Dichtl
  et~al\mbox{.}}{2017}]%
        {Dichtl2017DielectricAcid}
\bibfield{author}{\bibinfo{person}{Claudius Dichtl}, \bibinfo{person}{Pit
  Sippel}, {and} \bibinfo{person}{Stephan Krohns}.}
  \bibinfo{year}{2017}\natexlab{}.
\newblock \showarticletitle{{Dielectric Properties of 3D Printed Polylactic
  Acid}}.
\newblock \bibinfo{journal}{\emph{Advances in Materials Science and
  Engineering}}  \bibinfo{volume}{2017} (\bibinfo{year}{2017}).
\newblock
\showISSN{16878442}
\urldef\tempurl%
\url{https://doi.org/10.1155/2017/6913835}
\showDOI{\tempurl}


\bibitem[\protect\citeauthoryear{Flowers, Reyes, Ye, Kim, and Wiley}{Flowers
  et~al\mbox{.}}{2017}]%
        {Flowers20173DFilament}
\bibfield{author}{\bibinfo{person}{Patrick~F. Flowers},
  \bibinfo{person}{Christopher Reyes}, \bibinfo{person}{Shengrong Ye},
  \bibinfo{person}{Myung~Jun Kim}, {and} \bibinfo{person}{Benjamin~J. Wiley}.}
  \bibinfo{year}{2017}\natexlab{}.
\newblock \showarticletitle{{3D printing electronic components and circuits
  with conductive thermoplastic filament}}.
\newblock \bibinfo{journal}{\emph{Additive Manufacturing}}
  \bibinfo{volume}{18}, \bibinfo{number}{2017} (\bibinfo{year}{2017}),
  \bibinfo{pages}{156--163}.
\newblock
\showISSN{22148604}
\urldef\tempurl%
\url{https://doi.org/10.1016/j.addma.2017.10.002}
\showDOI{\tempurl}


\bibitem[\protect\citeauthoryear{Groeger and Steimle}{Groeger and
  Steimle}{2018}]%
        {10.1145/3161165}
\bibfield{author}{\bibinfo{person}{Daniel Groeger} {and}
  \bibinfo{person}{Jürgen Steimle}.} \bibinfo{year}{2018}\natexlab{}.
\newblock \showarticletitle{{ObjectSkin}}.
\newblock \bibinfo{journal}{\emph{Proceedings of the ACM on Interactive,
  Mobile, Wearable and Ubiquitous Technologies}} \bibinfo{volume}{1},
  \bibinfo{number}{4} (\bibinfo{date}{1} \bibinfo{year}{2018}),
  \bibinfo{pages}{1--23}.
\newblock
\showISSN{2474-9567}
\urldef\tempurl%
\url{https://doi.org/10.1145/3161165}
\showDOI{\tempurl}


\bibitem[\protect\citeauthoryear{Hanton, Wessely, Mueller, Fraser, and
  Roudaut}{Hanton et~al\mbox{.}}{2020}]%
        {10.1145/3334480.3383174}
\bibfield{author}{\bibinfo{person}{Ollie Hanton}, \bibinfo{person}{Michael
  Wessely}, \bibinfo{person}{Stefanie Mueller}, \bibinfo{person}{Mike Fraser},
  {and} \bibinfo{person}{Anne Roudaut}.} \bibinfo{year}{2020}\natexlab{}.
\newblock \showarticletitle{{ProtoSpray: Combining 3D printing and spraying to
  create interactive displays with arbitrary shapes}}. In
  \bibinfo{booktitle}{\emph{Conference on Human Factors in Computing Systems -
  Proceedings}} \emph{(\bibinfo{series}{CHI '20})}.
  \bibinfo{publisher}{Association for Computing Machinery},
  \bibinfo{address}{New York, NY, USA}, \bibinfo{pages}{1–4}.
\newblock
\showISBNx{9781450368193}
\urldef\tempurl%
\url{https://doi.org/10.1145/3334480.3383174}
\showDOI{\tempurl}


\bibitem[\protect\citeauthoryear{Huang, Wu, Xiao, Duan, Zhu, Bian, Ye, and
  Yin}{Huang et~al\mbox{.}}{2019}]%
        {Huang2019AssemblySurfaces}
\bibfield{author}{\bibinfo{person}{Yongan Huang}, \bibinfo{person}{Hao Wu},
  \bibinfo{person}{Lin Xiao}, \bibinfo{person}{Yongqing Duan},
  \bibinfo{person}{Hui Zhu}, \bibinfo{person}{Jing Bian}, \bibinfo{person}{Dong
  Ye}, {and} \bibinfo{person}{Zhouping Yin}.} \bibinfo{year}{2019}\natexlab{}.
\newblock \showarticletitle{{Assembly and applications of 3D conformal
  electronics on curvilinear surfaces}}.
\newblock \bibinfo{journal}{\emph{Materials Horizons}} \bibinfo{volume}{6},
  \bibinfo{number}{4} (\bibinfo{year}{2019}), \bibinfo{pages}{642--683}.
\newblock
\showISSN{20516355}
\urldef\tempurl%
\url{https://doi.org/10.1039/c8mh01450g}
\showDOI{\tempurl}


\bibitem[\protect\citeauthoryear{Jr, Billah, Carrasco, Barraza, Wicker, and
  Espalin}{Jr et~al\mbox{.}}{2018}]%
        {Jr2018HybridFabrication}
\bibfield{author}{\bibinfo{person}{Jose L~Coronel Jr},
  \bibinfo{person}{Kazi~Masum Billah}, \bibinfo{person}{Carlos F~Acosta
  Carrasco}, \bibinfo{person}{Sol~A Barraza}, \bibinfo{person}{Ryan~B Wicker},
  {and} \bibinfo{person}{David Espalin}.} \bibinfo{year}{2018}\natexlab{}.
\newblock \showarticletitle{{Hybrid Manufacturing with FDM Technology for
  Enabling Power Electronics Component Fabrication}}. In
  \bibinfo{booktitle}{\emph{Solid Freeform Fabrication 2018}}.
  \bibinfo{pages}{357--364}.
\newblock


\bibitem[\protect\citeauthoryear{Kim, Espalin, Liang, Xin, Cuaron, Varela,
  Macdonald, and Wicker}{Kim et~al\mbox{.}}{2017}]%
        {Kim2017}
\bibfield{author}{\bibinfo{person}{Chiyen Kim}, \bibinfo{person}{David
  Espalin}, \bibinfo{person}{Min Liang}, \bibinfo{person}{Hao Xin},
  \bibinfo{person}{Alejandro Cuaron}, \bibinfo{person}{Issac Varela},
  \bibinfo{person}{Eric Macdonald}, {and} \bibinfo{person}{Ryan~B. Wicker}.}
  \bibinfo{year}{2017}\natexlab{}.
\newblock \showarticletitle{{3D printed electronics with high performance,
  multi-layered electrical interconnect}}.
\newblock \bibinfo{journal}{\emph{IEEE Access}}  \bibinfo{volume}{5}
  (\bibinfo{year}{2017}), \bibinfo{pages}{25286--25294}.
\newblock
\showISSN{21693536}
\urldef\tempurl%
\url{https://doi.org/10.1109/ACCESS.2017.2773571}
\showDOI{\tempurl}


\bibitem[\protect\citeauthoryear{Kim, Cruz, Ye, Gray, Smith, Lazarus, Walker,
  Sigmarsson, and Wiley}{Kim et~al\mbox{.}}{2019}]%
        {Kim2019}
\bibfield{author}{\bibinfo{person}{Myung~Jun Kim}, \bibinfo{person}{Mutya~A.
  Cruz}, \bibinfo{person}{Shengrong Ye}, \bibinfo{person}{Allen~L. Gray},
  \bibinfo{person}{Gabriel~L. Smith}, \bibinfo{person}{Nathan Lazarus},
  \bibinfo{person}{Christopher~J. Walker}, \bibinfo{person}{Hjalti~H.
  Sigmarsson}, {and} \bibinfo{person}{Benjamin~J. Wiley}.}
  \bibinfo{year}{2019}\natexlab{}.
\newblock \showarticletitle{{One-step electrodeposition of copper on conductive
  3D printed objects}}.
\newblock \bibinfo{journal}{\emph{Additive Manufacturing}}
  \bibinfo{volume}{27}, \bibinfo{number}{March} (\bibinfo{year}{2019}),
  \bibinfo{pages}{318--326}.
\newblock
\showISSN{22148604}
\urldef\tempurl%
\url{https://doi.org/10.1016/j.addma.2019.03.016}
\showDOI{\tempurl}


\bibitem[\protect\citeauthoryear{Lazarus, Bedair, Hawasli, Kim, Wiley, and
  Smith}{Lazarus et~al\mbox{.}}{2019}]%
        {Lazarus2019}
\bibfield{author}{\bibinfo{person}{Nathan Lazarus}, \bibinfo{person}{Sarah~S.
  Bedair}, \bibinfo{person}{Sami~H. Hawasli}, \bibinfo{person}{Myung~Jun Kim},
  \bibinfo{person}{Benjamin~J. Wiley}, {and} \bibinfo{person}{Gabriel~L.
  Smith}.} \bibinfo{year}{2019}\natexlab{}.
\newblock \showarticletitle{{Selective Electroplating for 3D-Printed
  Electronics}}.
\newblock \bibinfo{journal}{\emph{Advanced Materials Technologies}}
  \bibinfo{volume}{4}, \bibinfo{number}{8} (\bibinfo{year}{2019}),
  \bibinfo{pages}{1--5}.
\newblock
\showISSN{2365709X}
\urldef\tempurl%
\url{https://doi.org/10.1002/admt.201900126}
\showDOI{\tempurl}


\bibitem[\protect\citeauthoryear{Lopes, MacDonald, and Wicker}{Lopes
  et~al\mbox{.}}{2012}]%
        {Lopes2012}
\bibfield{author}{\bibinfo{person}{Amit~Joe Lopes}, \bibinfo{person}{Eric
  MacDonald}, {and} \bibinfo{person}{Ryan~B. Wicker}.}
  \bibinfo{year}{2012}\natexlab{}.
\newblock \showarticletitle{{Integrating stereolithography and direct print
  technologies for 3D structural electronics fabrication}}.
\newblock \bibinfo{journal}{\emph{Rapid Prototyping Journal}}
  \bibinfo{volume}{18}, \bibinfo{number}{2} (\bibinfo{year}{2012}),
  \bibinfo{pages}{129--143}.
\newblock
\showISSN{13552546}
\urldef\tempurl%
\url{https://doi.org/10.1108/13552541211212113}
\showDOI{\tempurl}


\bibitem[\protect\citeauthoryear{Markvicka, Wang, Lee, Laput, Majidi, and
  Yao}{Markvicka et~al\mbox{.}}{2019}]%
        {10.1145/3290605.3300862}
\bibfield{author}{\bibinfo{person}{Eric Markvicka}, \bibinfo{person}{Guanyun
  Wang}, \bibinfo{person}{Yi-Chin Lee}, \bibinfo{person}{Gierad Laput},
  \bibinfo{person}{Carmel Majidi}, {and} \bibinfo{person}{Lining Yao}.}
  \bibinfo{year}{2019}\natexlab{}.
\newblock \showarticletitle{{ElectroDermis: Fully Untethered, Stretchable, and
  Highly-Customizable Electronic Bandages}}. In
  \bibinfo{booktitle}{\emph{Proceedings of the 2019 CHI Conference on Human
  Factors in Computing Systems}} \emph{(\bibinfo{series}{CHI '19})}.
  \bibinfo{publisher}{Association for Computing Machinery},
  \bibinfo{address}{New York, NY, USA}, \bibinfo{pages}{1–10}.
\newblock
\showISBNx{9781450359702}
\urldef\tempurl%
\url{https://doi.org/10.1145/3290605.3300862}
\showDOI{\tempurl}


\bibitem[\protect\citeauthoryear{Mueller, Kruck, and Baudisch}{Mueller
  et~al\mbox{.}}{2013}]%
        {Mueller2013LaserOrigami:Objects}
\bibfield{author}{\bibinfo{person}{Stefanie Mueller}, \bibinfo{person}{Bastian
  Kruck}, {and} \bibinfo{person}{Patrick Baudisch}.}
  \bibinfo{year}{2013}\natexlab{}.
\newblock \showarticletitle{{LaserOrigami: Laser-cutting 3D objects}}.
\newblock \bibinfo{journal}{\emph{Conference on Human Factors in Computing
  Systems - Proceedings}} (\bibinfo{year}{2013}), \bibinfo{pages}{2585--2592}.
\newblock
\showISBNx{9781450318990}
\urldef\tempurl%
\url{https://doi.org/10.1145/2470654.2481358}
\showDOI{\tempurl}


\bibitem[\protect\citeauthoryear{Oh, Ta, Suzuki, Gross, Kawahara, and Yao}{Oh
  et~al\mbox{.}}{2018}]%
        {10.1145/3173574.3174015}
\bibfield{author}{\bibinfo{person}{Hyunjoo Oh}, \bibinfo{person}{Tung~D Ta},
  \bibinfo{person}{Ryo Suzuki}, \bibinfo{person}{Mark~D Gross},
  \bibinfo{person}{Yoshihiro Kawahara}, {and} \bibinfo{person}{Lining Yao}.}
  \bibinfo{year}{2018}\natexlab{}.
\newblock \showarticletitle{{PEP (3D Printed Electronic Papercrafts): An
  Integrated Approach for 3D Sculpting Paper-Based Electronic Devices}}. In
  \bibinfo{booktitle}{\emph{Proceedings of the 2018 CHI Conference on Human
  Factors in Computing Systems}} \emph{(\bibinfo{series}{CHI '18})}.
  \bibinfo{publisher}{Association for Computing Machinery},
  \bibinfo{address}{New York, NY, USA}, \bibinfo{pages}{1–12}.
\newblock
\showISBNx{9781450356206}
\urldef\tempurl%
\url{https://doi.org/10.1145/3173574.3174015}
\showDOI{\tempurl}


\bibitem[\protect\citeauthoryear{Olberding, Ortega, Hildebrandt, and
  Steimle}{Olberding et~al\mbox{.}}{2015}]%
        {Olberding2015Foldio:Electronics}
\bibfield{author}{\bibinfo{person}{Simon Olberding},
  \bibinfo{person}{Sergio~Soto Ortega}, \bibinfo{person}{Klaus Hildebrandt},
  {and} \bibinfo{person}{Jürgen Steimle}.} \bibinfo{year}{2015}\natexlab{}.
\newblock \showarticletitle{{Foldio: Digital fabrication of interactive and
  shape-changing objects with foldable printed electronics}}.
\newblock \bibinfo{journal}{\emph{UIST 2015 - Proceedings of the 28th Annual
  ACM Symposium on User Interface Software and Technology}}
  (\bibinfo{year}{2015}), \bibinfo{pages}{223--232}.
\newblock
\showISBNx{9781450337793}
\urldef\tempurl%
\url{https://doi.org/10.1145/2807442.2807494}
\showDOI{\tempurl}


\bibitem[\protect\citeauthoryear{Roquet, Kim, and Yeh}{Roquet
  et~al\mbox{.}}{2016}]%
        {Roquet20163DCircuits}
\bibfield{author}{\bibinfo{person}{Claudia~Daudén Roquet},
  \bibinfo{person}{Jeeeun Kim}, {and} \bibinfo{person}{Tom Yeh}.}
  \bibinfo{year}{2016}\natexlab{}.
\newblock \showarticletitle{{3D folded PrintGami: Transform passive 3D printed
  objects to interactive by inserted paper Origami circuits}}.
\newblock \bibinfo{journal}{\emph{DIS 2016 - Proceedings of the 2016 ACM
  Conference on Designing Interactive Systems: Fuse}} (\bibinfo{year}{2016}),
  \bibinfo{pages}{187--191}.
\newblock
\showISBNx{9781450340311}
\urldef\tempurl%
\url{https://doi.org/10.1145/2901790.2901891}
\showDOI{\tempurl}


\bibitem[\protect\citeauthoryear{Saada, Layani, Chernevousky, and
  Magdassi}{Saada et~al\mbox{.}}{2017}]%
        {Saada2017HydroprintingStructures}
\bibfield{author}{\bibinfo{person}{Gabriel Saada}, \bibinfo{person}{Michael
  Layani}, \bibinfo{person}{Avi Chernevousky}, {and} \bibinfo{person}{Shlomo
  Magdassi}.} \bibinfo{year}{2017}\natexlab{}.
\newblock \showarticletitle{{Hydroprinting Conductive Patterns onto 3D
  Structures}}.
\newblock \bibinfo{journal}{\emph{Advanced Materials Technologies}}
  \bibinfo{volume}{2}, \bibinfo{number}{5} (\bibinfo{year}{2017}),
  \bibinfo{pages}{1--6}.
\newblock
\showISSN{2365709X}
\urldef\tempurl%
\url{https://doi.org/10.1002/admt.201600289}
\showDOI{\tempurl}


\bibitem[\protect\citeauthoryear{Savage, Zhang, and Hartmann}{Savage
  et~al\mbox{.}}{2012}]%
        {Savage2012Midas:Objects}
\bibfield{author}{\bibinfo{person}{Valkyrie Savage}, \bibinfo{person}{Xiaohan
  Zhang}, {and} \bibinfo{person}{Björn Hartmann}.}
  \bibinfo{year}{2012}\natexlab{}.
\newblock \showarticletitle{{Midas: Fabricating custom capacitive touch sensors
  to prototype interactive objects}}.
\newblock \bibinfo{journal}{\emph{UIST'12 - Proceedings of the 25th Annual ACM
  Symposium on User Interface Software and Technology}} (\bibinfo{year}{2012}),
  \bibinfo{pages}{579--587}.
\newblock
\showISBNx{9781450315807}


\bibitem[\protect\citeauthoryear{Schmitz, Khalilbeigi, Balwierz, Lissermann,
  M{\"{u}}hlha{\"{u}}ser, and Steimle}{Schmitz et~al\mbox{.}}{2015}]%
        {Schmitz2015Capricate:Objects}
\bibfield{author}{\bibinfo{person}{Martin Schmitz},
  \bibinfo{person}{Mohammadreza Khalilbeigi}, \bibinfo{person}{Matthias
  Balwierz}, \bibinfo{person}{Roman Lissermann}, \bibinfo{person}{Max
  M{\"{u}}hlha{\"{u}}ser}, {and} \bibinfo{person}{Jühlhürgen Steimle}.}
  \bibinfo{year}{2015}\natexlab{}.
\newblock \showarticletitle{{Capricate: A fabrication pipeline to design and 3D
  printcapacitive touch sensors for interactive objects}}.
\newblock \bibinfo{journal}{\emph{UIST 2015 - Proceedings of the 28th Annual
  ACM Symposium on User Interface Software and Technology}}
  (\bibinfo{year}{2015}), \bibinfo{pages}{253--258}.
\newblock
\showISBNx{9781450337793}
\urldef\tempurl%
\url{https://doi.org/10.1145/2807442.2807503}
\showDOI{\tempurl}


\bibitem[\protect\citeauthoryear{Schmitz, Stitz, M{\"{u}}ller, Funk, and
  M{\"{u}}hlh{\"{a}}user}{Schmitz et~al\mbox{.}}{2019}]%
        {10.1145/3290605.3300684}
\bibfield{author}{\bibinfo{person}{Martin Schmitz}, \bibinfo{person}{Martin
  Stitz}, \bibinfo{person}{Florian M{\"{u}}ller}, \bibinfo{person}{Markus
  Funk}, {and} \bibinfo{person}{Max M{\"{u}}hlh{\"{a}}user}.}
  \bibinfo{year}{2019}\natexlab{}.
\newblock \showarticletitle{{../Trilaterate: A Fabrication Pipeline to Design
  and 3D Print Hover-, Touch-, and Force-Sensitive Objects}}. In
  \bibinfo{booktitle}{\emph{Proceedings of the 2019 CHI Conference on Human
  Factors in Computing Systems}} \emph{(\bibinfo{series}{CHI '19})}.
  \bibinfo{publisher}{Association for Computing Machinery},
  \bibinfo{address}{New York, NY, USA}, \bibinfo{pages}{1–13}.
\newblock
\showISBNx{9781450359702}
\urldef\tempurl%
\url{https://doi.org/10.1145/3290605.3300684}
\showDOI{\tempurl}


\bibitem[\protect\citeauthoryear{Sch{\"{u}}ller, Panozzo, Grundh{\"{o}}fer,
  Zimmer, Sorkine, and Sorkine-Hornung}{Sch{\"{u}}ller et~al\mbox{.}}{2016}]%
        {Schuller2016ComputationalThermoforming}
\bibfield{author}{\bibinfo{person}{Christian Sch{\"{u}}ller},
  \bibinfo{person}{Daniele Panozzo}, \bibinfo{person}{Anselm Grundh{\"{o}}fer},
  \bibinfo{person}{Henning Zimmer}, \bibinfo{person}{Evgeni Sorkine}, {and}
  \bibinfo{person}{Olga Sorkine-Hornung}.} \bibinfo{year}{2016}\natexlab{}.
\newblock \showarticletitle{{Computational thermoforming}}.
\newblock \bibinfo{journal}{\emph{ACM Transactions on Graphics}}
  \bibinfo{volume}{35}, \bibinfo{number}{4} (\bibinfo{year}{2016}),
  \bibinfo{pages}{2--10}.
\newblock
\showISBNx{9781450342797}
\showISSN{15577368}
\urldef\tempurl%
\url{https://doi.org/10.1145/2897824.2925914}
\showDOI{\tempurl}


\bibitem[\protect\citeauthoryear{Swaminathan, Ozutemiz, Majidi, and
  Hudson}{Swaminathan et~al\mbox{.}}{2019}]%
        {10.1145/3290605.3300797}
\bibfield{author}{\bibinfo{person}{Saiganesh Swaminathan},
  \bibinfo{person}{Kadri~Bugra Ozutemiz}, \bibinfo{person}{Carmel Majidi},
  {and} \bibinfo{person}{Scott~E. Hudson}.} \bibinfo{year}{2019}\natexlab{}.
\newblock \showarticletitle{{FiberWire}}. In
  \bibinfo{booktitle}{\emph{Proceedings of the 2019 CHI Conference on Human
  Factors in Computing Systems}} \emph{(\bibinfo{series}{CHI '19})}.
  \bibinfo{publisher}{Association for Computing Machinery},
  \bibinfo{address}{New York, NY, USA}, \bibinfo{pages}{1--11}.
\newblock
\showISBNx{9781450359702}
\urldef\tempurl%
\url{https://doi.org/10.1145/3290605.3300797}
\showDOI{\tempurl}


\bibitem[\protect\citeauthoryear{Takada, Shizuki, and Tanaka}{Takada
  et~al\mbox{.}}{2016}]%
        {Takada2016MonoTouch:Gestures}
\bibfield{author}{\bibinfo{person}{Ryosuke Takada}, \bibinfo{person}{Buntarou
  Shizuki}, {and} \bibinfo{person}{Jiro Tanaka}.}
  \bibinfo{year}{2016}\natexlab{}.
\newblock \showarticletitle{{MonoTouch: Single capacitive touch sensor that
  differentiates touch gestures}}.
\newblock \bibinfo{journal}{\emph{Conference on Human Factors in Computing
  Systems - Proceedings}}  \bibinfo{volume}{07-12-May-} (\bibinfo{year}{2016}),
  \bibinfo{pages}{2736--2743}.
\newblock
\showISBNx{9781450340823}
\urldef\tempurl%
\url{https://doi.org/10.1145/2851581.2892350}
\showDOI{\tempurl}


\bibitem[\protect\citeauthoryear{Tino, Moore, Antoline, Ravi, Wake, Ionita,
  Morris, Decker, Sheikh, Rybicki, and Chepelev}{Tino et~al\mbox{.}}{2020}]%
        {Tino2020COVID-19Medicine}
\bibfield{author}{\bibinfo{person}{Rance Tino}, \bibinfo{person}{Ryan Moore},
  \bibinfo{person}{Sam Antoline}, \bibinfo{person}{Prashanth Ravi},
  \bibinfo{person}{Nicole Wake}, \bibinfo{person}{Ciprian~N. Ionita},
  \bibinfo{person}{Jonathan~M. Morris}, \bibinfo{person}{Summer~J. Decker},
  \bibinfo{person}{Adnan Sheikh}, \bibinfo{person}{Frank~J. Rybicki}, {and}
  \bibinfo{person}{Leonid~L. Chepelev}.} \bibinfo{year}{2020}\natexlab{}.
\newblock \showarticletitle{{COVID-19 and the role of 3D printing in
  medicine}}.
\newblock \bibinfo{journal}{\emph{3D Printing in Medicine}}
  \bibinfo{volume}{6}, \bibinfo{number}{1} (\bibinfo{year}{2020}),
  \bibinfo{pages}{1--8}.
\newblock
\showISSN{2365-6271}
\urldef\tempurl%
\url{https://doi.org/10.1186/s41205-020-00064-7}
\showDOI{\tempurl}


\bibitem[\protect\citeauthoryear{Umetani and Schmidt}{Umetani and
  Schmidt}{2017}]%
        {Umetani2017SurfCuit:Prints}
\bibfield{author}{\bibinfo{person}{Nobuyuki Umetani} {and}
  \bibinfo{person}{Ryan Schmidt}.} \bibinfo{year}{2017}\natexlab{}.
\newblock \showarticletitle{{SurfCuit: Surface-Mounted Circuits on 3D Prints}}.
\newblock \bibinfo{journal}{\emph{IEEE Computer Graphics and Applications}}
  \bibinfo{volume}{38}, \bibinfo{number}{3} (\bibinfo{year}{2017}),
  \bibinfo{pages}{52--60}.
\newblock
\showISSN{02721716}
\urldef\tempurl%
\url{https://doi.org/10.1109/MCG.2017.40}
\showDOI{\tempurl}


\bibitem[\protect\citeauthoryear{Van{\v{e}}{\v{c}}kov{\'{a}}, Bou{\v{s}}a,
  Sokolov{\'{a}}, Moreno-Garc{\'{i}}a, Broekmann, Shestivska, Rathousk{\'{y}},
  G{\'{a}}l, Sebechlebsk{\'{a}}, and
  Kolivo{\v{s}}ka}{Van{\v{e}}{\v{c}}kov{\'{a}} et~al\mbox{.}}{2020}]%
        {Vaneckova2020CopperElectrodes}
\bibfield{author}{\bibinfo{person}{Eva Van{\v{e}}{\v{c}}kov{\'{a}}},
  \bibinfo{person}{Milan Bou{\v{s}}a}, \bibinfo{person}{Romana Sokolov{\'{a}}},
  \bibinfo{person}{Pavel Moreno-Garc{\'{i}}a}, \bibinfo{person}{Peter
  Broekmann}, \bibinfo{person}{Violetta Shestivska}, \bibinfo{person}{Jiří
  Rathousk{\'{y}}}, \bibinfo{person}{Miroslav G{\'{a}}l},
  \bibinfo{person}{Táňa Sebechlebsk{\'{a}}}, {and} \bibinfo{person}{Viliam
  Kolivo{\v{s}}ka}.} \bibinfo{year}{2020}\natexlab{}.
\newblock \showarticletitle{{Copper electroplating of 3D printed composite
  electrodes}}.
\newblock \bibinfo{journal}{\emph{Journal of Electroanalytical Chemistry}}
  \bibinfo{volume}{858} (\bibinfo{year}{2020}).
\newblock
\showISSN{15726657}
\urldef\tempurl%
\url{https://doi.org/10.1016/j.jelechem.2019.113763}
\showDOI{\tempurl}


\bibitem[\protect\citeauthoryear{Varun~Perumal and Wigdor}{Varun~Perumal and
  Wigdor}{2015}]%
        {10.1145/2807442.2807511}
\bibfield{author}{\bibinfo{person}{C. Varun~Perumal} {and}
  \bibinfo{person}{Daniel Wigdor}.} \bibinfo{year}{2015}\natexlab{}.
\newblock \showarticletitle{{Printem: Instant printed circuit boards with
  standard office printers and inks}}. In \bibinfo{booktitle}{\emph{UIST 2015 -
  Proceedings of the 28th Annual ACM Symposium on User Interface Software and
  Technology}} \emph{(\bibinfo{series}{UIST '15})}.
  \bibinfo{publisher}{Association for Computing Machinery},
  \bibinfo{address}{New York, NY, USA}, \bibinfo{pages}{243--251}.
\newblock
\showISBNx{9781450337793}
\urldef\tempurl%
\url{https://doi.org/10.1145/2807442.2807511}
\showDOI{\tempurl}


\bibitem[\protect\citeauthoryear{Vatani, Engeberg, and Choi}{Vatani
  et~al\mbox{.}}{2015a}]%
        {Vatani2015ConformalSensors}
\bibfield{author}{\bibinfo{person}{Morteza Vatani}, \bibinfo{person}{Erik~D.
  Engeberg}, {and} \bibinfo{person}{Jae~Won Choi}.}
  \bibinfo{year}{2015}\natexlab{a}.
\newblock \showarticletitle{{Conformal direct-print of piezoresistive
  polymer/nanocomposites for compliant multi-layer tactile sensors}}.
\newblock \bibinfo{journal}{\emph{Additive Manufacturing}}  \bibinfo{volume}{7}
  (\bibinfo{year}{2015}), \bibinfo{pages}{73--82}.
\newblock
\showISSN{22148604}
\urldef\tempurl%
\url{https://doi.org/10.1016/j.addma.2014.12.009}
\showDOI{\tempurl}


\bibitem[\protect\citeauthoryear{Vatani, Lu, Engeberg, and Choi}{Vatani
  et~al\mbox{.}}{2015b}]%
        {Vatani2015CombinedSensors}
\bibfield{author}{\bibinfo{person}{Morteza Vatani}, \bibinfo{person}{Yanfeng
  Lu}, \bibinfo{person}{Erik~D. Engeberg}, {and} \bibinfo{person}{Jae~Won
  Choi}.} \bibinfo{year}{2015}\natexlab{b}.
\newblock \showarticletitle{{Combined 3D printing technologies and material for
  fabrication of tactile sensors}}.
\newblock \bibinfo{journal}{\emph{International Journal of Precision
  Engineering and Manufacturing}} \bibinfo{volume}{16}, \bibinfo{number}{7}
  (\bibinfo{year}{2015}), \bibinfo{pages}{1375--1383}.
\newblock
\showISSN{20054602}
\urldef\tempurl%
\url{https://doi.org/10.1007/s12541-015-0181-3}
\showDOI{\tempurl}


\bibitem[\protect\citeauthoryear{Wang, Huo, Chawla, Chen, Banerjee, and
  Ramani}{Wang et~al\mbox{.}}{2018}]%
        {Wang2018Plain2Fun:Circuits}
\bibfield{author}{\bibinfo{person}{Tianyi Wang}, \bibinfo{person}{Ke Huo},
  \bibinfo{person}{Pratik Chawla}, \bibinfo{person}{Guiming Chen},
  \bibinfo{person}{Siddharth Banerjee}, {and} \bibinfo{person}{Karthik
  Ramani}.} \bibinfo{year}{2018}\natexlab{}.
\newblock \showarticletitle{{Plain2Fun: Augmenting ordinary objects with
  interactive functions by auto-fabricating surface painted circuits}}.
\newblock \bibinfo{journal}{\emph{DIS 2018 - Proceedings of the 2018 Designing
  Interactive Systems Conference}} (\bibinfo{year}{2018}),
  \bibinfo{pages}{1095--1106}.
\newblock
\showISBNx{9781450351980}
\urldef\tempurl%
\url{https://doi.org/10.1145/3196709.3196791}
\showDOI{\tempurl}


\bibitem[\protect\citeauthoryear{Wasserfall}{Wasserfall}{2015}]%
        {Wasserfall2015}
\bibfield{author}{\bibinfo{person}{Florens Wasserfall}.}
  \bibinfo{year}{2015}\natexlab{}.
\newblock \showarticletitle{{Embedding of SMD populated circuits into FDM
  printed objects}}. In \bibinfo{booktitle}{\emph{SFF Symposium Proceedings}}.
  \bibinfo{pages}{180--189}.
\newblock


\bibitem[\protect\citeauthoryear{Wu, Yang, Hsu, and Lin}{Wu
  et~al\mbox{.}}{2015}]%
        {Wu20153D-printedSensors}
\bibfield{author}{\bibinfo{person}{Sung~Yueh Wu}, \bibinfo{person}{Chen Yang},
  \bibinfo{person}{Wensyang Hsu}, {and} \bibinfo{person}{Liwei Lin}.}
  \bibinfo{year}{2015}\natexlab{}.
\newblock \showarticletitle{{3D-printed microelectronics for integrated
  circuitry and passive wireless sensors}}.
\newblock \bibinfo{journal}{\emph{Microsystems and Nanoengineering}}
  \bibinfo{volume}{1}, \bibinfo{number}{June} (\bibinfo{year}{2015}),
  \bibinfo{pages}{1--9}.
\newblock
\showISSN{20557434}
\urldef\tempurl%
\url{https://doi.org/10.1038/micronano.2015.13}
\showDOI{\tempurl}


\bibitem[\protect\citeauthoryear{Yamaoka, Dogan, Bulovic, Saito, Kawahara,
  Kakehi, and Mueller}{Yamaoka et~al\mbox{.}}{2019}]%
        {10.1145/3290605.3300858}
\bibfield{author}{\bibinfo{person}{Junichi Yamaoka},
  \bibinfo{person}{Mustafa~Doga Dogan}, \bibinfo{person}{Katarina Bulovic},
  \bibinfo{person}{Kazuya Saito}, \bibinfo{person}{Yoshihiro Kawahara},
  \bibinfo{person}{Yasuaki Kakehi}, {and} \bibinfo{person}{Stefanie Mueller}.}
  \bibinfo{year}{2019}\natexlab{}.
\newblock \showarticletitle{{FoldTronics: Creating 3D objects with integrated
  electronics using foldable honeycomb structures}}. In
  \bibinfo{booktitle}{\emph{Conference on Human Factors in Computing Systems -
  Proceedings}} \emph{(\bibinfo{series}{CHI '19})}.
  \bibinfo{publisher}{Association for Computing Machinery},
  \bibinfo{address}{New York, NY, USA}, \bibinfo{pages}{1–14}.
\newblock
\showISBNx{9781450359702}
\urldef\tempurl%
\url{https://doi.org/10.1145/3290605.3300858}
\showDOI{\tempurl}


\bibitem[\protect\citeauthoryear{Zhu, Blumberg, Zhu, Nisser, Carlson, Wen,
  Shum, Quaye, and Mueller}{Zhu et~al\mbox{.}}{2020a}]%
        {10.1145/3313831.3376617}
\bibfield{author}{\bibinfo{person}{Junyi Zhu}, \bibinfo{person}{Lotta~Gili
  Blumberg}, \bibinfo{person}{Yunyi Zhu}, \bibinfo{person}{Martin Nisser},
  \bibinfo{person}{Ethan~Levi Carlson}, \bibinfo{person}{Xin Wen},
  \bibinfo{person}{Kevin Shum}, \bibinfo{person}{Jessica~Ayeley Quaye}, {and}
  \bibinfo{person}{Stefanie Mueller}.} \bibinfo{year}{2020}\natexlab{a}.
\newblock \showarticletitle{{CurveBoards demo: Integrating breadboards into
  physical objects to prototype function in the context of form}}.
\newblock In \bibinfo{booktitle}{\emph{Conference on Human Factors in Computing
  Systems - Proceedings}}. \bibinfo{publisher}{Association for Computing
  Machinery}, \bibinfo{address}{New York, NY, USA}, \bibinfo{pages}{1–13}.
\newblock
\showISBNx{9781450368193}
\urldef\tempurl%
\url{https://doi.org/10.1145/3334480.3383149}
\showDOI{\tempurl}


\bibitem[\protect\citeauthoryear{Zhu, Zhu, Cui, Cheng, Snowden, Chounlakone,
  Wessely, and Mueller}{Zhu et~al\mbox{.}}{2020b}]%
        {10.1145/3379337.3415898}
\bibfield{author}{\bibinfo{person}{Junyi Zhu}, \bibinfo{person}{Yunyi Zhu},
  \bibinfo{person}{Jiaming Cui}, \bibinfo{person}{Leon Cheng},
  \bibinfo{person}{Jackson Snowden}, \bibinfo{person}{Mark Chounlakone},
  \bibinfo{person}{Michael Wessely}, {and} \bibinfo{person}{Stefanie Mueller}.}
  \bibinfo{year}{2020}\natexlab{b}.
\newblock \showarticletitle{{MorphSensor: A 3D electronic design tool for
  reforming sensor modules}}. In \bibinfo{booktitle}{\emph{UIST 2020 -
  Proceedings of the 33rd Annual ACM Symposium on User Interface Software and
  Technology}} \emph{(\bibinfo{series}{UIST '20})}.
  \bibinfo{publisher}{Association for Computing Machinery},
  \bibinfo{address}{New York, NY, USA}, \bibinfo{pages}{541--553}.
\newblock
\showISBNx{9781450375146}
\urldef\tempurl%
\url{https://doi.org/10.1145/3379337.3415898}
\showDOI{\tempurl}


\end{thebibliography}


\end{document}